\documentstyle[12pt,epsfig]{article}
\def\bge{\begin{equation}}
\def\ene{\end{equation}}
\def\bg{\begin{eqnarray}}
\def\en{\end{eqnarray}}
\def\nn{\nonumber}
\def\vecp{\vec{p}}
\begin{document}
%%%%%%%%%%%%%%%%%%%%%%%%%%%%%%%%%%%%%%%%%%%%%%%%%%%%%%%%%%%%%%%%%%%%%%
\begin{titlepage}
\title{Resonance model study of kaon production in baryon baryon
reactions for heavy ion collisions}
\author{
K. Tsushima$^1$~\thanks{ktsushim@physics.adelaide.edu.au} ,
A. Sibirtsev$^2$~\thanks{sibirt@theorie.physik.uni-giessen.de 
and on leave from the Institute of Theoretical and Experimental Physics, 
117259 Moscow, Russia} ,
A. W. Thomas$^1$~\thanks{athomas@physics.adelaide.edu.au} , 
G. Q. Li$^3$ ~\thanks{gqli@nulcear.physics.sunysb.edu}\\
{\small $^1$Department of Physics and Mathematical Physics } \\
{\small and Special Research Center for the Subatomic 
Structure of Matter} \\
{\small University of Adelaide, SA 5005, Australia} \\
{$^2$\small Institut f\"ur Theoretische Physik, Universit\"at Giessen} \\
{\small D-35392 Giessen, Germany}  \\
{$^3$\small Department of Physics and Astronomy, State University of
New York at Stony Brook } \\
{\small Stony Brook, New York 11794, USA}} 
\date{}
\maketitle
\vspace{-10.5cm}
\hfill ADP-98-1/T282,\quad UGI-98-3,\quad SB-NTG-98-32
\vspace{10.5cm}
\begin{abstract}
The energy dependence of the total kaon production cross sections 
in baryon baryon ($N$ and $\Delta$) collisions are studied 
in the resonance model, which is a relativistic, tree-level treatment.
This study is the first attempt to complete   
a systematic, consistent investigation of the elementary kaon
production reactions for both the pion baryon and baryon baryon reactions.
Our model suggests that the magnitudes of the isospin-averaged total 
cross sections for the $N N \to N Y K$ and $\Delta N \to N Y K$
($Y = \Lambda$ or $\Sigma$) reactions are almost equal 
at energies up to about 200 MeV above threshold.
However, the magnitudes for the $\Delta N$ reactions
become about 6 times larger than those for the $N N$ reactions
at energies about 1 GeV above threshold. Furthermore, the magnitudes
of the isospin-averaged total cross sections for the 
$N N \to \Delta Y K$ reactions turn out to be comparable to those for 
the $N N \to N Y K$ reactions at $N N$ invariant collision energies about  
3.1 GeV, and about 5 to 10 times larger at $N N$ invariant 
collision energies about 3.5 GeV. The microscopic cross sections
are parametrized in all isospin channels necessary for the transport model
studies of kaon production in heavy ion collisions. These cross
sections are then applied in the relativistic transport model
to study the sensitivity to the underlying elementary kaon
production cross sections.
\end{abstract}
\end{titlepage}
%%%%%%%%%%%%%%%%%%%%%%%%%%%%%%%%%%%%%%%%%%%%%%%%%%%%%%%%%%%%%%%%%%%%%%
%
%Section 1

\section{Introduction}

Particle production in heavy ion collisions is a unique tool for studying
the properties of matter under extreme conditions such as
high temperature and/or density~\cite{cas1}-\cite{gro}. 
In particular, positive kaon, $K^+$, production in heavy ion collisions 
at intermediate energies is one of the most promising 
methods to probe matter formed under such conditions in the central zone
of the collisions~\cite{sch,nag}. 

Because $K^+$-mesons have a long mean free path due to the  
small cross sections for scattering on nucleons, they can 
provide us almost undisturbed information about the central zone of 
heavy ion collisions, in which most of the $K^+$-mesons are considered 
to be produced in subthreshold heavy ion reactions.
According to the theoretical investigations in Refs.~\cite{aic1}-\cite{li1}, 
the kaon yield is also very sensitive to 
the nuclear equation of state. 
Furthermore, kaon production in heavy ion collisions may  
provide clues concerning quark gluon plasma formation
from the $K/\pi$ production ratio~\cite{pkoc,abb}, 
chiral symmetry restoration~\cite{bro,vkoc,brho},
and the existence of density isomers~\cite{har1}. 
Thus, many studies of kaon production in heavy ion collisions 
have been made experimentally and theoretically~\cite{sto}-\cite{lilast}.

The emphasis and focus of earlier theoretical studies was 
on the momentum-dependent nuclear interactions~\cite{aic2,li3,maru},
effects of the rescattering and kaon mass  
renormalization~\cite{fan}, effects of chiral symmetry 
restoration~\cite{bro}, and the medium modification of elementary
kaon production cross sections~\cite{wu,fanko}. However, most of these  
theoretical investigations were perfomed using the energy dependence of 
the total kaon production cross sections parametrized in a simple 
phenomenological way~\cite{schu,ran,cug1}. As a consequence, all these 
investigations involve a certain amount of ambiguity arising 
from such parametrizations.
Furthermore, although recent investigations 
of strangeness production emphasize the importance of 
elementary processes such as the 
$B_1 B_2 \to B_3 Y K$ and $\pi B \to Y K$ reactions~\cite{lilee,kaos}, 
no investigations for these reactions have been perfomed in a consistent 
manner even in a tree-level treatment. 
(Here $B_{1,2,3}$ and $B$ stand for either the 
nucleon or the $\Delta$, while $Y$ stands for either the $\Lambda$
or the $\Sigma$ hyperon.)

Although many issues need to be considered in order to achieve a better  
understanding of kaon production in heavy ion collisions, 
the one emphasized in this article is the 
elementary kaon production cross sections at the hadronic 
level as was studied in 
Refs.~\cite{ferrari,yao,deloff,laget,sib2,libbk,tsuppk,tsu}. 
Since most of the microscopic transport models assume that 
the relevant reactions are described by on-shell, binary interactions
at the hadronic level, it should be straightforward  
to improve the parametrizations for the elementary kaon production 
cross sections, which are important inputs of the models.  

Recently, the COSY-11 Collaboration~\cite{cosy} measured the  
total cross section for the $p p \to p \Lambda K^+$ reaction 
at an energy 2 MeV above the reaction threshold.
The result distinctly differs from those of the commonly
used phenomenological parametrizations~\cite{schu,ran}.
However, in practice, this discrepancy probably does not significantly   
influence the past results calculated using the  
parametrizations made without this new data point, because the 
magnitudes of the cross sections at this energy are very 
small for both the experiment~\cite{cosy} 
and parametrizations~\cite{schu,ran}.

On the other hand, theoretical reanalyses in Refs.~\cite{Cassingn,fuc,debow}  
using different parametrizations~\cite{sib2} for  
the $N N \to N Y K$ cross sections, indicate that 
the secondary reactions, $\pi N \to Y K$, give rather important  
contributions -- comparable with those of the baryon baryon ($N$ and $\Delta$) 
induced reactions. This conclusion contradicts the previously
accepted scenario, in which the dominant contributions for $K^+$ 
production in heavy ion collisions come from $\Delta N$ and 
$\Delta \Delta$ collisions~\cite{lan,hua2,har2,bali,libbk}.
Thus, it is necessary to investigate further the elementary kaon 
production cross sections for both the $\pi B \to Y K$ and
$B_1 B_2 \to B_3 Y K$ reactions in a consistent manner.
Furthermore, all the conclusions drawn about the elementary reactions 
involving a $\Delta$ were based on the results obtained using 
the elementary cross sections for the nucleon, not for the $\Delta$, 
except for the isospin difference.
Thus, an explicit calculation for the cross sections  
involving a $\Delta$ should be performed, treating properly not 
only isospin degrees of freedom but also spin as was initially  
attempted in Ref.~\cite{libbk} using the results of the resonance 
model for the $\pi B \to Y K$ reactions~\cite{tsu}.

For this purpose, we complete our  
systematic investigations for the elementary kaon production 
processes in all baryon baryon reactions, extending the studies 
made so far for the pion baryon and proton proton reactions 
using the resonance model~\cite{tsuppk,tsulambda}.
The treatment of the present study is a tree-level, using empirical 
branching ratios for the relevant resonances. It is relativistic, 
and incorporates the pion baryon, 
and the baryon baryon reactions in a consistent manner.
Thus, most of the cut-off parameters, 
coupling constants and form factors have been fixed 
already in pion baryon and proton proton reactions~\cite{tsuppk,tsu}.
Furthermore, we parametrize the energy dependence of these  
total kaon production cross sections in baryon baryon collisions
for all the isospin channels needed in the simulation codes.
We note that the total cross sections for the  
$p p \to p \Lambda K^+$ reaction recently measured at energies 
about 50 and 150 MeV above threshold~\cite{bil} 
show an excellent agreement 
with the predicted results of the resonance model~\cite{tsuppk}. 
However, because the treatment in the model does not include the 
final state interactions, our results cannot be expected to describe 
well the near-threshold behaviour.
A more rigorous treatment of the resonances consistent with 
the unitary condition was studied by Fuester and Mosel~\cite{fuester}, 
but such an approach does not seem to be suitable for the present purpose, 
and beyond the scope of the present study.

The organization of this article is as follows.  
In Sect.~2 we will explain the resonance model in detail. 
Effective Lagrangian densities at the hadronic level together with 
the experimental data for the model will be described.
Numerical results for the total cross sections 
will be presented in Sec.~3. In Sect.~4 we give all the 
parametrizations for the energy dependence of the total cross sections 
calculated in the model. 
In Sec. 5 these cross sections are used in the relativistic
transport model to study the sensitivity of the transport
model predictions to the underlying elementary kaon
production cross sections.
Discussions and summary will be given in Sect.~6. 
Finally, in the Appendix, the 
relations between the branching ratios of the resonances and 
the coupling constants relevant for the model will be given.
In addition, we will supplement the cross section relations 
for the $\pi N \to \Lambda K$ reaction which were not mentioned 
in Ref.~\cite{tsu}. 

%%%%%%%%%%%%%%%%%%%%%%%%%%%%%%%%%%%%%%%%%%%%%%%%%%%%%%%%%%%%%%%%%%%%%%%%%%%%
%Section 2
\section{Resonance model}

\begin{figure}[hbt]
\epsfig{file=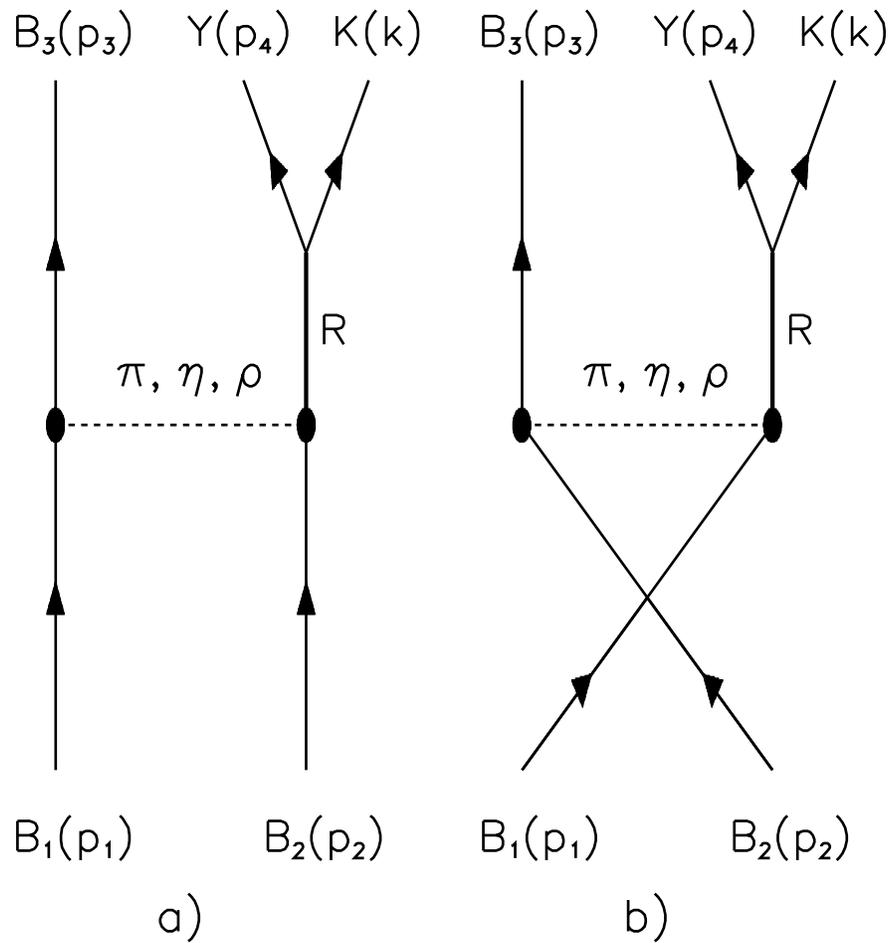,width=16cm}
\vspace{-1cm}
\caption{Kaon production processes in the resonance model.
$B_i\, (i=1,2,3)$, $Y$ and $R$ stand for respectively,  
either the nucleon or the $\Delta$, either the $\Lambda$ or the $\Sigma$ 
hyperon, and the baryon resonances.}
\label{diagram}
\end{figure}

In Fig.~\ref{diagram}, we show the Feynman diagrams relevant 
for kaon production in the resonance model.
$R$ stands for the baryon resonances which are responsible 
for the kaon and hyperon pair production,  
with their masses up to about 2 GeV.
Assumptions and approximations of the model are~\cite{tsuppk,tsulambda,tsu}: 
\begin{enumerate}
\item Resonances which are experimentally observed to 
decay to hyperon and kaon are included in the model. 
A kaon is always produced from these resonances simultaneously
with a hyperon. In the study of the $\pi B \to Y K$ reactions, we have also 
investigated the non-resonant contributions, 
including an effective $t$-channel 
$K^*$-meson exchange. It turned out that this non-resonant 
contribution was small in order to reproduce the experimental data, 
once the relevant $s$-channel resonances were included in the 
model~\cite{tsu}.
Thus, kaon exchange which is included in 
Refs.~\cite{ferrari,deloff,laget,sib2,libbk} and 
would introduce extra new parameters in the present treatment,  
is not included in the present calculation.
Instead, other meson exchanges, namely $\eta$ and $\rho$ meson 
exchanges, are included, 
and they compensate the contribution of kaon 
exchange introduced in other models.
\item Mesons exchanged are restricted to those
observed in the decay channels of the adopted resonances. 
(See also Fig.~\ref{diagram}.)
We note that there is a discussion in Ref.~\cite{tsulambda} 
of whether there is a possibility to  
settle what kind of meson exchange is responsible 
for the $p p \to p \Lambda K^+$ reaction.
\item Those processes in which the exchanged pion can be on-shell are 
excluded, because on-shell pions are usually included as  
secondary processes in microscopic transport models through 
processes such as $\pi N \leftrightarrow \Delta$ and 
$\pi B \to Y K$. Thus, the $\Delta \Delta \to N Y K$ reactions, 
which are possible only through $\pi$-exchange in the model,
are not included. Note that because of the assumption made in number 2,  
neither $\rho$-meson nor $\eta$-meson exchanges will give contributions
to this channel in the present treatment.
\item Resonances are treated as elementary particles, except that their  
observed widths enter in the propagators. Because of this,
$\rho$-exchange attached to the $N(1650)$ resonance is excluded 
kinematically, although this resonance has a branch in the 
$N \rho$ channel~\cite{par}.
\item Any final state interactions which are usually important for 
describing the near-threshold behaviour are not included. 
Thus, our results should not be taken seriously near the 
reaction threshold, which is different from the present main purpose.
\end{enumerate}
It is worth noting that all values of the coupling constants squared
relevant for the meson-baryon-(baryon resonance) vertices,
$g^2_{M B R}$ $(M = \pi, \eta, \rho)$, can be determined from the
experimental decay ratios, once the interaction Lagrangian densities 
and relevant form factors are specified. 
In Table~\ref{resonance} we summarize the data for those 
resonances which are included in the model, and necessary to calculate 
the coupling constants.

%%%%%%%%%%%%%%%%%%%%%%%%%%%%%%%%%%%%%%%%%%%%%%%%%%%%%%%%%%%%%%%%%%%%%%%%%%%%
\begin{table}
\caption{\label{resonance}
Resonances included in the model.
Confidence levels of the resonances are,
$N(1650)****$, $N(1710)***$, $N(1720)****$ and 
$\Delta(1920)***$~\protect\cite{par}. Note that 
the $\Delta(1920)$ resonance is treated as an effective 
resonance which represents all contributions of six resonances,  
$\Delta(1900)$, $\Delta(1905)$, $\Delta(1910)$, $\Delta(1920)$, 
$\Delta(1930)$ and $\Delta(1940)$. 
See Ref.~\protect\cite{tsu} for this effective treatment 
of the $\Delta(1920)$.}
\begin{center}
\begin{tabular}{|c|cccc|}
\hline
Resonance $(J^P)$ &Width (MeV) &Decay channel &Branching ratio &Adopted value \\
\hline
N(1650)$\,(\frac{1}{2}^-)$ &150 &$N \pi$      &0.60 -- 0.80 &0.700 \\
                           &    &$N \eta$     &0.03 -- 0.10 &0.065 \\
                           &    &$\Delta \pi$ &0.03 -- 0.07 &0.050 \\
                           &    &$\Lambda K$  &0.03 -- 0.11 &0.070 \\
\hline
N(1710)$\,(\frac{1}{2}^+)$ &100 &$N \pi$      &0.10 -- 0.20 &0.150 \\
                           &    &$N \eta$     &0.20 -- 0.40 &0.300 \\
                           &    &$N \rho$     &0.05 -- 0.25 &0.150 \\
                           &    &$\Delta \pi$ &0.10 -- 0.25 &0.175 \\
                           &    &$\Lambda K$  &0.05 -- 0.25 &0.150 \\
                           &    &$\Sigma K$   &0.02 -- 0.10 &0.060 \\
\hline
N(1720)$\,(\frac{3}{2}^+)$ &150 &$N \pi$      &0.10 -- 0.20 &0.150 \\
                           &    &$N \eta$     &0.02 -- 0.06 &0.040 \\
                           &    &$N \rho$     &0.70 -- 0.85 &0.775 \\
                           &    &$\Delta \pi$ &0.05 -- 0.15 &0.100 \\
                           &    &$\Lambda K$  &0.03 -- 0.10 &0.065 \\
                           &    &$\Sigma K$   &0.02 -- 0.05 &0.035 \\
\hline
$\Delta$(1920)$\,(\frac{3}{2}^+)$ &200 &$N \pi$ &0.05 -- 0.20 &0.125 \\
                           &    &$\Sigma K$   &0.01 -- 0.03 &0.020 \\
\hline
\end{tabular}
\end{center}
\end{table}
%%%%%%%%%%%%%%%%%%%%%%%%%%%%%%%%%%%%%%%%%%%%%%%%%%%%%%%%%%%%%%%%%%%%%%%%%%%%

Effective Lagrangian densities relevant for evaluating the
Feynman diagrams depicted in Fig.~\ref{diagram} are: 
%%%%%%%%%%%%%%%%%%%%%%%%%%%%%%%%%%%%%%
\begin{eqnarray}
{\cal L}_{\pi N N} &=& 
-ig_{\pi N N} \bar{N} \gamma_5 \vec\tau N \cdot \vec\pi,
\label{pnn}\\
{\cal L}_{\pi N \Delta} &=&  
- \frac{g_{\pi N \Delta}}{m_\pi} 
\left( \bar{\Delta}^\mu \overrightarrow{\cal I} N \cdot 
\partial_\mu \vec\pi + \bar{N} {\overrightarrow{\cal I}}^\dagger 
\Delta^\mu \cdot \partial_\mu \vec\pi \, \right),
\label{pni}\\
{\cal L}_{\pi N N(1650)} &=&  
-g_{\pi N N(1650)}
\left( \bar{N}(1650) \vec\tau N \cdot \vec\pi
+ \bar{N} \vec\tau N(1650) \cdot \vec\pi\,\, \right), 
\label{pna}\\
{\cal L}_{\pi N N(1710)} &=&  
-ig_{\pi N N(1710)}
\left( \bar{N}(1710) \gamma_5 \vec\tau N \cdot \vec\pi
+ \bar{N} \gamma_5 \vec\tau N(1710) \cdot \vec\pi\,\, \right), 
\label{pnb}\\
{\cal L}_{\pi N N(1720)} &=&
\frac{g_{\pi N N(1720)}}{m_\pi}
\left( \bar{N}^\mu(1720) \vec\tau N \cdot \partial_\mu \vec\pi
+ \bar{N} \vec\tau N^\mu(1720) \cdot \partial_\mu \vec\pi \, \right),
\label{pnc}\\
{\cal L}_{\pi N \Delta(1920)} &=& 
\frac{g_{\pi N \Delta(1920)}}{m_\pi}
\left( \bar{\Delta}^\mu(1920) \overrightarrow{\cal I} N \cdot 
\partial_\mu \vec\pi + \bar{N} {\overrightarrow{\cal I}}^\dagger 
\Delta^\mu(1920) \cdot \partial_\mu \vec\pi \, \right),
\label{pnd}\\
{\cal L}_{\pi \Delta \Delta} &=& 
- i g_{\pi \Delta \Delta} \bar{\Delta}_\mu \gamma_5 
\vec{\cal K} \Delta^\mu \cdot \vec\pi,
\label{pii}\\
{\cal L}_{\pi \Delta N(1650)} &=&  
i \frac{g_{\pi \Delta N(1650)}}{m_\pi} 
\left( \bar{N}(1650) \gamma_5 {\overrightarrow{\cal I}}^\dagger 
\Delta^\mu \cdot \partial_\mu \vec\pi
+ \bar{\Delta}^\mu \gamma_5 \overrightarrow{\cal I} N(1650) 
\cdot \partial_\mu \vec\pi \, \right),
\label{pia}\\
{\cal L}_{\pi \Delta N(1710)} &=&  
\frac{g_{\pi \Delta N(1710)}}{m_\pi} 
\left( \bar{N}(1710) {\overrightarrow{\cal I}}^\dagger 
\Delta^\mu \cdot \partial_\mu \vec\pi
+ \bar{\Delta} \overrightarrow{\cal I} N(1710) 
\cdot \partial_\mu \vec\pi \, \right),
\label{pib}\\
{\cal L}_{\pi \Delta N(1720)} &=&  
- i g_{\pi \Delta N(1720)} 
\left( \bar{N}_\mu(1720) \gamma_5 {\overrightarrow{\cal I}}^\dagger 
\Delta^\mu \cdot \vec\pi
+ \bar{\Delta}^\mu \gamma_5 \overrightarrow{\cal I} N^\mu (1720) 
\cdot \vec\pi \, \right),
\label{pic}\\
{\cal L}_{\eta N N} &=& 
-ig_{\eta N N} \bar{N} \gamma_5 N \eta,
\label{enn}\\
{\cal L}_{\eta \Delta \Delta} &=& 
-ig_{\eta \Delta \Delta} \bar{\Delta}_\mu \gamma_5 \Delta^\mu \eta,
\label{eii}\\
{\cal L}_{\eta N N(1650)} &=&
- g_{\eta N N(1650)}
\left( \bar{N}(1650) N \eta
+ \bar{N} N(1650) \eta\,\, \right),
\label{ena}\\
{\cal L}_{\eta N N(1710)} &=&
-ig_{\eta N N(1710)}
\left( \bar{N}(1710) \gamma_5 N \eta
+ \bar{N} \gamma_5 N(1710) \eta\,\, \right),
\label{enb}\\
{\cal L}_{\eta N N(1720)} &=&
\frac{g_{\eta N N(1720)}}{m_\eta}
\left( \bar{N}^\mu(1720) N \partial_\mu \eta
+ \bar{N} N^\mu(1720) \partial_\mu \eta \, \right),
\label{enc}\\
{\cal L}_{\rho N N} &=& - g_{\rho N N}
\left( \bar{N} \gamma^\mu \vec\tau N \cdot \vec\rho_\mu
+ \frac{\kappa}{2 m_N} \bar{N} \sigma^{\mu \nu}
\vec\tau N \cdot \partial_\mu \vec\rho_\nu \right),
\label{rnn}\\
{\cal L}_{\rho \Delta \Delta} &=& 
g_{\rho \Delta \Delta} \bar{\Delta}_\mu \gamma^\nu 
\vec{\cal K} \Delta^\mu \cdot \vec\rho_\nu,
\label{rii}\\
{\cal L}_{\rho N \Delta} &=&  
- i \frac{g_{\rho N \Delta}}{m_\rho} 
\left( \bar{\Delta}^\mu \gamma^\nu \gamma_5 \overrightarrow{\cal I} N 
+ \bar{N} \gamma^\mu \gamma_5 
{\overrightarrow{\cal I}}^\dagger \Delta^\nu \right) 
\cdot 
( \partial_\mu \vec\rho_\nu - \partial_\nu \vec\rho_\mu ), 
\label{rni}\\
{\cal L}_{\rho N N(1710)} &=&
-g_{\rho N N(1710)}
\left( \bar{N}(1710) \gamma^\mu \vec\tau N \cdot \vec\rho_\mu
+ \bar{N} \gamma^\mu \vec\tau N(1710) \cdot \vec\rho_\mu\,\, \right),
\label{rnb}\\
{\cal L}_{\rho N N(1720)} &=&
-ig_{\rho N N(1720)}
\left( \bar{N}^\mu(1720) \gamma_5 \vec\tau N \cdot \vec\rho_\mu
+ \bar{N} \gamma_5 \vec\tau N^\mu(1720) \cdot \vec\rho_\mu\,\, \right),
\label{rnc}\\
{\cal L}_{K \Lambda N(1650)} &=&
-g_{K \Lambda N(1650)}
\left( \bar{N}(1650) \Lambda K + \bar{K} \bar\Lambda  N(1650) \right),
\label{kla}\\
{\cal L}_{K \Lambda N(1710)} &=&
-ig_{K \Lambda N(1710)}
\left( \bar{N}(1710) \gamma_5 \Lambda K
+ \bar{K} \bar\Lambda \gamma_5 N(1710) \right),
\label{klb}\\
{\cal L}_{K \Lambda N(1720)} &=&
\frac{g_{K \Lambda N(1720)}}{m_K}
\left( \bar{N}^\mu(1720) \Lambda
\partial_\mu K + (\partial_\mu \bar{K}) \bar\Lambda N^\mu(1720) \right),
\label{klc}\\
{\cal L}_{K \Sigma N(1710)} &=&
-ig_{K \Sigma N(1710)}
\left( \bar{N}(1710) \gamma_5 \vec\tau \cdot \overrightarrow\Sigma K
+ \bar{K} \overrightarrow{\bar \Sigma} \cdot \vec\tau
\gamma_5 N(1710) \right),
\label{ksb}\\
{\cal L}_{K \Sigma N(1720)} &=&
\frac{g_{K \Sigma N(1720)}}{m_K}
\left( \bar{N}^\mu(1720) \vec\tau \cdot \overrightarrow\Sigma
\partial_\mu K + (\partial_\mu \bar{K}) \overrightarrow{\bar \Sigma}
\cdot \vec\tau N^\mu(1720) \right),
\label{ksc}\\
{\cal L}_{K \Sigma \Delta(1920)} &=&
\frac{g_{K \Sigma \Delta(1920)}}{m_K}
\left( \bar{\Delta}^\mu(1920) \overrightarrow{\cal I}
\cdot \overrightarrow\Sigma \partial_\mu K
+ (\partial_\mu \bar{K}) \overrightarrow{\bar \Sigma} \cdot
{\overrightarrow{\cal I}}^\dagger \Delta^\mu(1920) \right).
\label{ksd}
\end{eqnarray}
%%%%%%%%%%%%%%%%%%%%%%%%%%%%%%%%%%%%%%
%
In the above, the operators $\overrightarrow{\cal I}$ 
and $\overrightarrow{\cal K}$ are defined by
\begin{eqnarray}
\overrightarrow{\cal I}_{M\mu} &\equiv&  \displaystyle{\sum_{\ell=\pm1,0}}
(1 \ell \frac{1}{2} \mu | \frac{3}{2} M)
\hat{e}^*_{\ell},\\ 
\overrightarrow{\cal K}_{M M'} &\equiv&  \displaystyle{\sum_{\ell=\pm1,0}}
(1 \ell \frac{3}{2} M' | \frac{3}{2} M)
\hat{e}^*_{\ell},\label{ccdelta}
\end{eqnarray}
with $M$, $\mu$ and $M'$ being the third components of the 
isospin projections, 
and $\vec \tau$ the Pauli matrices.
$N, N(1710), N(1720)$ and $\Delta(1920)$
stand for the fields of the nucleon
and the corresponding baryon resonances. They are expressed by
$\bar{N} = \left( \bar{p}, \bar{n} \right)$,
similarly for the nucleon resonances, and
$\bar{\Delta}(1920) = ( \bar{\Delta}(1920)^{++},
\bar{\Delta}(1920)^+, \bar{\Delta}(1920)^0,
\bar{\Delta}(1920)^- )$ in isospin space.
The physical representations of the kaon field are, 
$K^T = \left( K^+, K^0 \right)$ and 
$\bar{K} = \left( K^-, \bar{K^0} \right)$, respectively, 
%
%\pi^{\pm} =  (\pi_1 \mp i \pi_2)/\sqrt{2},\,
%\pi^0 = \pi_3,\,\,$ similarly for the $\rho$-meson fields, and 
%
%$\Sigma^{\pm} = (\Sigma_1 \mp i \Sigma_2)/{\sqrt{2}},\,\,
%\Sigma^0 = \Sigma_3\,\,$, respectively, 
%
where the superscript $T$ means the 
transposition operation.
They are defined as annihilating (creating) the physical particle
(anti-particle) states.
For the propagators $iS_F(p)$ of the spin 1/2 and
$iG^{\mu \nu}(p)$ of the spin 3/2 resonances we use:
\begin{equation}
iS_F(p) = i \frac{\gamma \cdot p + m}{p^2 - m^2 + im\Gamma^{full}}\,,
\label{spin1/2}
\end{equation}
\begin{equation}
iG^{\mu \nu}(p) = i \frac{-P^{\mu \nu}(p)}{p^2 - m^2 +
im\Gamma^{full}}\,,  \label{spin3/2}
\end{equation}
with
\begin{equation}
P^{\mu \nu}(p) = - (\gamma \cdot p + m)
\left[ g^{\mu \nu} - \frac{1}{3} \gamma^\mu \gamma^\nu
- \frac{1}{3 m}( \gamma^\mu p^\nu - \gamma^\nu p^\mu)
- \frac{2}{3 m^2} p^\mu p^\nu \right], \label{pmunu}
\end{equation}
where $m$ and $\Gamma^{full}$ stand for the mass and full decay
width of the corresponding resonances.
For the form factors, $F_M (\vec{q})$ 
($\vec{q}$ is the momentum of meson, $M$), 
appearing in the meson-baryon-(baryon resonance) vertices, we use
\begin{equation}
F_M(\vec{q}) = 
\displaystyle{\left( \frac{\Lambda_M^2}{\Lambda_M^2 
+ \vec{q}\,^2} \right)^n}, 
\label{formfactor}
\end{equation}
where $n = 1$ for the $\pi$ and $\eta$-mesons and $n = 2$ for  
the $\rho$-meson, respectively, with $\Lambda_M$ being the cut-off parameter.
Most of the form factors, coupling constants
and cut-off parameters for the relevant meson-baryon-baryon and 
meson-baryon-(baryon resonance)  
vertices are adopted from Refs.~\cite{tsuppk,tsu}.
In the Appendix, we give the relations between the branching
ratios and the corresponding coupling constants squared,
$g^2_{M B R}$, which were calculated using the relevant Lagrangian  
densities. In the calculation, the form factors of
Eq. (\ref{formfactor}) are multiplied by the corresponding
coupling constants, $g_{M B R}$.
For the coupling constants, $g_{M \Delta \Delta}$ $(M = \pi, \eta, \rho)$,
which appeared for the first time in the present study,
we use an SU(6) quark model result with the definition Eq. (\ref{ccdelta}),
$g_{M \Delta \Delta} = 3 g_{M N N}$. In addition, we use the same value 
for the corresponding cut-off parameter as that for the nucleon.
For the value of the cut-off parameter at the $\rho N \Delta$ vertex, 
we use $544 = 1300 \times (920/2200)$ MeV, 
which is scaled the same amount as was necessary for 
the $\rho N N$ vertex of the Bonn potential model 
(the Model I in Table~B.1~\cite{mac}).
The other quantities, coupling constants, cut-off
parameters and form factors are, as far as possible, taken from the
same version of the model~\cite{mac}.
In addition, we use a value, $\kappa = f_{\rho N N}/g_{\rho N N} = 6.1$,
for the tensor coupling constant at the $\rho N N$ vertex.
We summarize in Table~\ref{cconst} all values for the coupling 
constants and cut-off parameters used in the study. 
%

%%%%%%%%%%%%%%%%%%%%%%%%%%%%%%%%%%%%%%%%%%%%%%%%%%
%Table (coupling constants)
\begin{table}
\caption{\label{cconst}
Coupling constants and cut-off parameters used in the present study. 
$\kappa = f_{\rho N N}/g_{\rho N N} = 6.1$ is used 
for the $\rho N N$ tensor coupling. Note that the coupling 
constants relevant for $\Delta(1920)$, $g_{\pi N \Delta(1920)}$ and 
$g_{K \Sigma \Delta(1920)}$, are scaled multiplying by a factor 
1.86 according to the effective treatment~\protect\cite{tsu}.  
See also the caption of Table~\protect\ref{resonance}. }
\begin{center}
\begin{tabular}{|l|l|c|c|l|l|c|}
\hline
vertex & $g^2/4\pi$ & cut-off (MeV)& 
&vertex &$g^2/4\pi$ &cut-off (MeV)\\
\hline 
 & & & & & & \\
 $\pi N N$  & $14.4$ & $1050$ & 
&$\pi \Delta \Delta$ &$3^2 \times 14.4$ &$1050$\\
 $\pi N N(1650)$ & $1.12 \times 10^{-1}$ & $800$ &
&$\pi N N(1710)$ &$2.05 \times 10^{-1}$ &$800$ \\
 $\pi N N(1720)$ &$4.13 \times 10^{-3}$ &$800$ &
&$\pi N \Delta(1920)$ &$1.13 \times 10^{-1}$ &$500$ \\
 $\pi \Delta N(1650)$ &$7.19 \times 10^{-2}$ &$800$ & 
&$\pi \Delta N(1710)$ &$2.23 \times 10^{-3}$ &$800$\\
 $\pi \Delta N(1720)$ &$1.39$ &$800$ &
&$\eta N N$ &$5.00$ &$2000$\\
 $\eta \Delta \Delta$ &$3^2 \times 5.00$ &$2000$ &
&$\eta N N(1650)$ &$3.37 \times 10^{-2}$ &$800$\\ 
 $\eta N N(1710)$ &$2.31$ &$800$ & 
&$\eta N N(1720)$ &$1.03 \times 10^{-1}$ &$800$\\
 $\rho N N$  &$0.74$ &$920$ & 
&$\rho N \Delta$ &$19.0$ &$544$ \\
 $\rho \Delta \Delta$  &$3^2 \times 0.74$ &$920$ & 
&$\rho N N(1710)$ &$3.61 \times 10^{+1}$ &$800$\\
 $\rho N N(1720)$  &$1.43 \times 10^{+2}$ &$800$ & 
&$K \Lambda N(1650)$ &$5.10 \times 10^{-2}$ &$800$\\
 $K \Lambda N(1710)$ &$3.78$ &$800$ &
&$K \Lambda N(1720)$ &$3.12 \times 10^{-1}$ &$800$\\
 $K \Sigma N(1710)$ &$4.66$ &$800$ &
&$K \Sigma N(1720)$ &$2.99 \times 10^{-1}$ &$800$\\
 $K \Sigma \Delta(1920)$ & $3.08 \times 10^{-1}$ & $500$ &
 & & &\\ 
 & & & & & & \\
\hline
\end{tabular}
\end{center}
\end{table}
%%%%%%%%%%%%%%%%%%%%%%%%%%%%%%%%%

%%%%%%%%%%%%%%%%%%%%%%%%%%%%%%%%%%%%%%%%%%%%%%%%%%%%%%%%%%%%%%%%%%%%%%
%Section 
\section{Numerical results}

In this study, we neglected all interference terms  
between the amplitudes. Thus, the relations given in this section 
for the cross sections are not always valid when
the interference terms are included rigorously.

\subsection{$ NN \to N Y K$}

Recently, the total cross section for 
the $p p \to p \Lambda K^+$ reaction was 
measured by the COSY 11 Collaboration~\cite{cosy} at energies 
of a few MeV above the reaction threshold, and it gave a new 
constraint on the theoretical calculations and phenomenological 
parametrizations. However, we do not refit the parameters and coupling 
constants to this new data point for the following reasons.
First, our treatment does not include the final state interactions 
nor interference terms, 
which will be important at energies very near threshold.
Second, the total cross sections at energies up to about 10~MeV 
above the threshold are very small in magnitude for both the parametrization 
and the new data point. Thus, this new data point will not influence 
the calculation of kaon yield in heavy ion collisions
(for instance see Fig.~9 in Ref.~\cite{Cassingn}).

Here, we write down explicitly the amplitude and 
cross section formula for the $p p \to p \Lambda K^+$ reaction,  
as an example. The total amplitude for this reaction is given by
\begin{eqnarray}
{\cal M}  &=& 
{\cal M}(p(1650), \pi^0) + {\cal M}(p(1710), \pi^0) + {\cal M}(p(1720), \pi^0)
\nonumber \\
&+& {\cal M}(p(1650),\eta) + {\cal M}(p(1710),\eta) + {\cal M}(p(1720),\eta)
\nonumber \\
& + & {\cal M}(p(1710),\rho^0)  +  {\cal M}(p(1720),\rho^0)
+ ({\rm exchange\,\, terms }),
\label{amplitude}
\end{eqnarray}
where on the right hand side of Eq.~(\ref{amplitude}), the resonances 
and mesons exchanged in the intermediate states are written
inside the bracket explicitly.
Each amplitude can be obtained straightforwardly by applying the  
Feynman rules with the relevant interaction Lagrangian densities.

For a given invariant collision energy, $\sqrt{s}$, the total cross 
section, $\sigma (p p \to p \Lambda K^+)$, can be calculated by
\begin{equation}
\sigma (p p \to p \Lambda K^+)
= \frac{1}{F} \int \, |\overline{{\cal M}}|^2 \, 
\delta^4(p_1+p_2-p_3-p_\Lambda-p_K) \,
\frac{d^3 p_3}{2 E_3} \, \frac{d^3 p_\Lambda}{2 E_\Lambda} 
\, \frac{d^3 p_K}{2 E_K},
\label{crosssection}
\end{equation}
with the flux factor 
\begin{eqnarray}
F &=& 2 \ \lambda^{1/2}(s,m_p^2,m_p^2) \ (2\pi)^5, \nonumber\\
\lambda(x,y,z) &\equiv& x^2 + y^2 + z^2 - 2xy - 2yz - 2zx,\nonumber
\end{eqnarray}
where the subscripts 1 and 2 stand for the initial protons, 
and 3 for the final proton, corresponding to the diagrams 
in Fig.~\ref{diagram} for $\sigma (B_1 B_2 \to B_3 Y K)$.
This notation, corresponding to Fig.~\ref{diagram}, will be used hereafter.
$|\overline{{\cal M}}|^2$ is the square of the scattering amplitude, 
averaged over the initial spins and summed over the final spins. 
To perform the integration over the final state 
three-body phase space, 
we first integrate on $d^3 p_3$ and $d^3 p_\Lambda$ 
in the center-of-mass (c.m.) frame of the final state proton and $\Lambda$,
and then integrate on $d^3 p_K$ in the c.m. frame 
of the initial protons~\cite{col} using the Lorentz transformation.
Note that our treatment neglects all interference terms and final state 
interactions~\cite{wilkin,alexfinal} which are important in 
near-threshold energy region. For the effect of the intereference terms
in the model, one can find a discussion in Ref.~\cite{tsuppk}. 

\begin{figure}[h]
\epsfig{figure=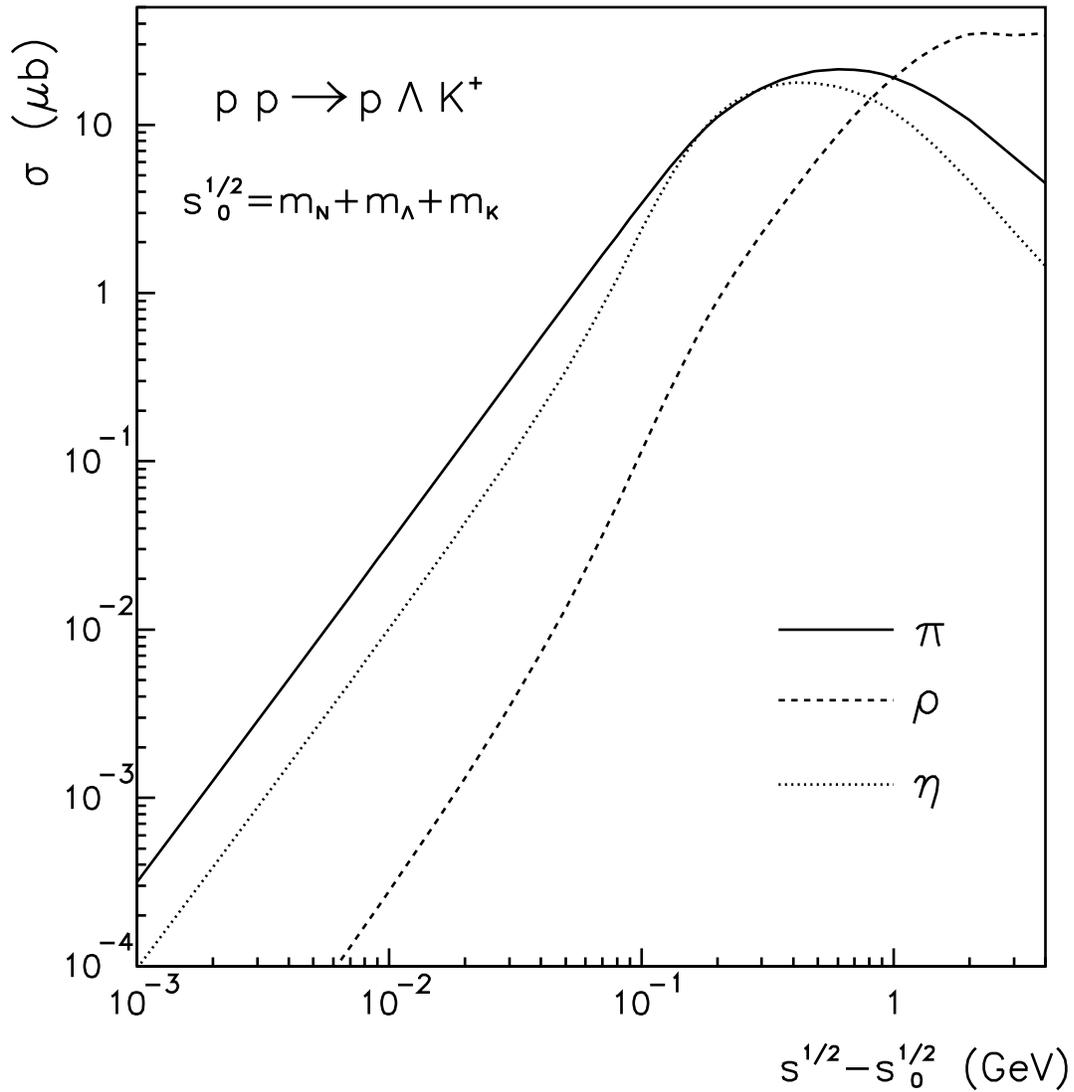,width=16cm}
\caption{\label{kt15p} Separate contributions from $\pi$, $\rho$
and $\eta$-exchanges to the total cross 
section for the $p p \to p \Lambda  K^+$ reaction. $s^{1/2}$ and 
$s^{1/2}_0 = m_N + m_\Lambda + m_K$ are the invariant collision 
energy and threshold energy, respectively.}
\end{figure}

Fig.~\ref{kt15p} illustrates the separate contributions from  
$\pi$, $\rho$ and $\eta$-exchanges to the total cross sections   
for the $p p \to p \Lambda K^+$ reaction. 
$\sqrt{s}$ is the $p p$ invariant collision energy
in their c.m. system, while 
$\sqrt{s_0}=m_N+m_\Lambda+m_K$ is the threshold energy 
with $m_N$, $m_\Lambda$ and $m_K$ being respectively 
the masses of the nucleon, $\Lambda$ and kaon.
One can see that pion exchange is dominant at energies near  
the reaction threshold, while $\rho$-exchange is dominant 
at higher energies.

\begin{figure}[hbt]
\epsfig{figure=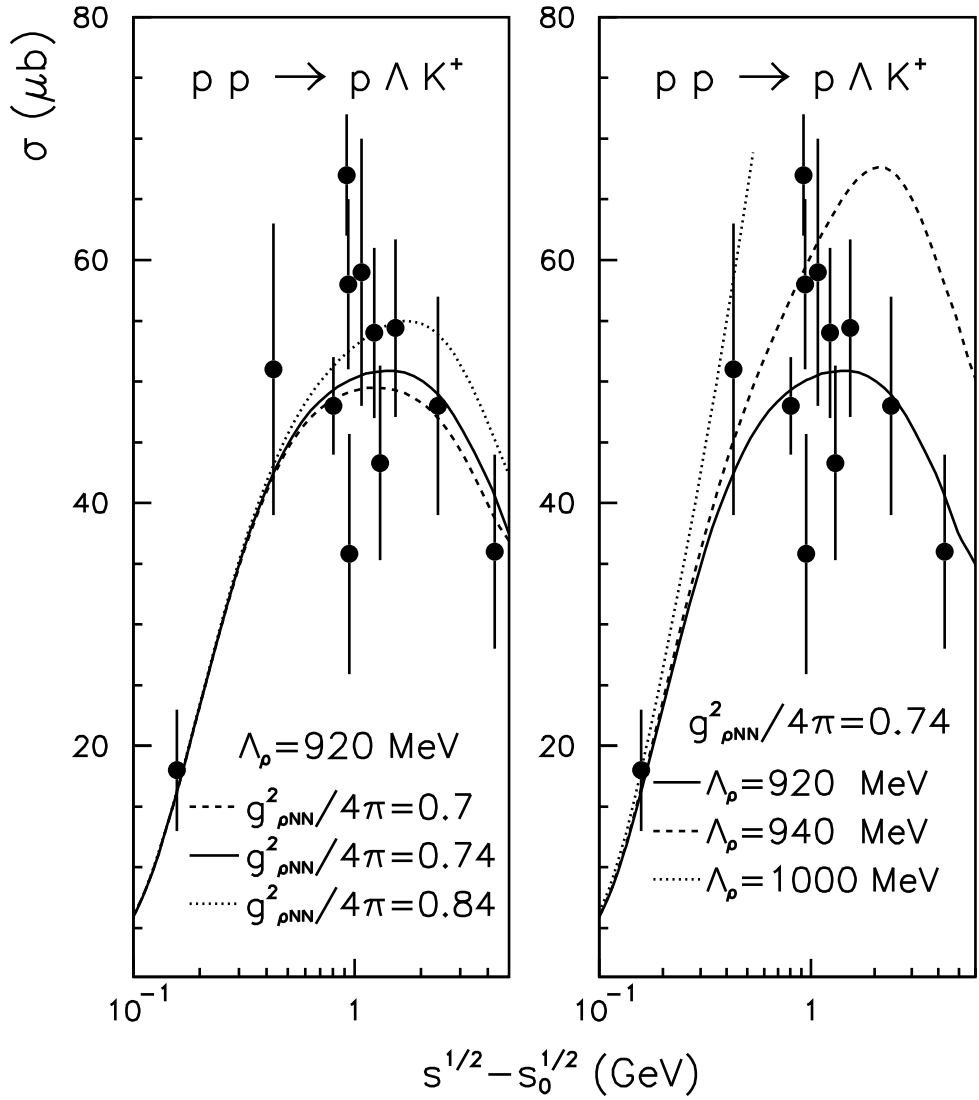,width=16cm}
\caption{\label{kt1p} Dependence on the $\rho NN$ coupling constant 
and cut-off parameter of the total cross section for
the $p p \to p \Lambda  K^+$ reaction.
The dots show experimental data from Ref.~\protect\cite{LB}.}
\end{figure}

As was already discussed in Ref.~\cite{tsuppk}, the total cross section 
is rather sensitive to the value of the cut-off parameter, $\Lambda_\rho$,  
in the $\rho N N$ vertex form factor at higher energies.
The dependence on the values of the cut-off parameter, $\Lambda_\rho$, and 
coupling constant, $g_{\rho N N}$, is  shown in Fig.~\ref{kt1p}.
After fixing the cut-off parameter   
to a specific value, the sensitivity of the total
cross section to the $\rho N N$
coupling constant, $g_{\rho N N}$,  
(with $\kappa$ = 6.1 for the tensor coupling constant) is small.
The best values to reproduce the experimental data~\cite{LB} at
energies larger than about 100 MeV above threshold were found 
to be, $g^2_{\rho N N}/4 \pi = 0.74$ and $\Lambda_\rho = 920$ MeV.

Since the total cross sections for the reaction at energies just 
above threshold receive their dominant contribution from pion exchange, 
we will test the sensitivity of the results to the value of the 
cut-off parameter in the $\pi N N$ form factor, $\Lambda_{\pi N N}$.  
Although the cut-off and coupling constant, 
$g_{\pi N N}$ and $\Lambda_{\pi N N}$, are not independent in a strict 
sense, when they are determined from the $N N$ phase shift data, 
we will show the results calculated using two different values for 
the cut-off parameter, $\Lambda_{\pi N N} = 3000$ MeV and $1050$ MeV, 
with the fixed, usually accepted value, 
$g^2_{\pi N N}/4\pi = 14.4$. The adopted value from this reaction 
for the cut-off parameter is, $\Lambda_{\pi N N} = 1050$ MeV.

\begin{figure}[hbt]
\epsfig{figure=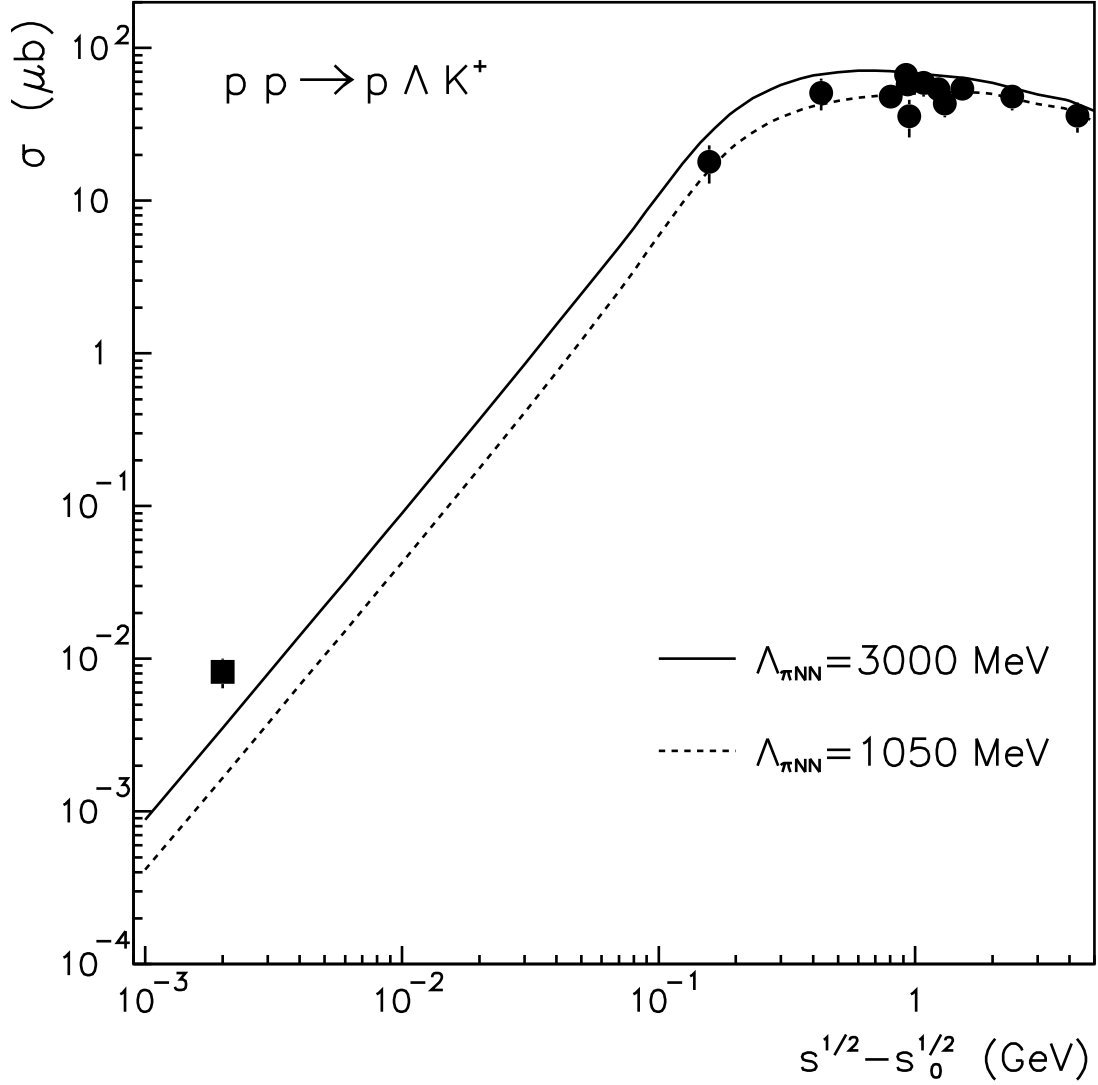,width=16cm}
\caption{\label{kt4p} Dependence on the cut-off parameter in the 
$\pi N N$ form factor, $\Lambda_{\pi N N}$, of the total cross sections  
for the $p p \to p \Lambda  K^+$ reaction.
The dots and square are the experimental data respectively from 
Ref.~\protect\cite{LB} and Ref.~\protect\cite{cosy}.}
\end{figure}

In Fig.~\ref{kt4p} we show the total cross sections for the
$p p \to p \Lambda K^+$ reaction calculated with the two different 
cut-offs, $\Lambda_{\pi N N} = 3000$ MeV (the solid line) and 
$\Lambda_{\pi N N} = 1050$ MeV (the dashed line), together with 
the experimental data~\cite{cosy,LB}.
Results are rather insensitive to the value, $\Lambda_{\pi N N}$. 
The new data point from COSY~\cite{cosy} does not seem 
to be achieved by varying the value of the $\Lambda_{\pi N N}$ 
within a reasonable range. However, it  
is worth noting that the parameter set fixed by the
$p p \to p \Lambda K^+$ reaction, including the cut-off $\Lambda_{\pi N N}$,  
must be used also for the calculation of the 
$N N \to N \Sigma K$ reactions.
It is not trivial at all if the same parameter set could also  
reproduce the experimental data for 
the $N N \to N \Sigma K$ reactions.    
Furthermore, although the parameters were optimised so as to reproduce 
the data points around 1 GeV above threshold, the  
recent data for the total cross sections measured at energies 
about 50 and 150 MeV above the threshold~\cite{bil} show an excellent 
agreement with the predicted values of the resonance model~\cite{tsuppk}.
This provides some confirmation of the validity of the parameters 
determined in the model.

\begin{figure}[h]
\epsfig{figure=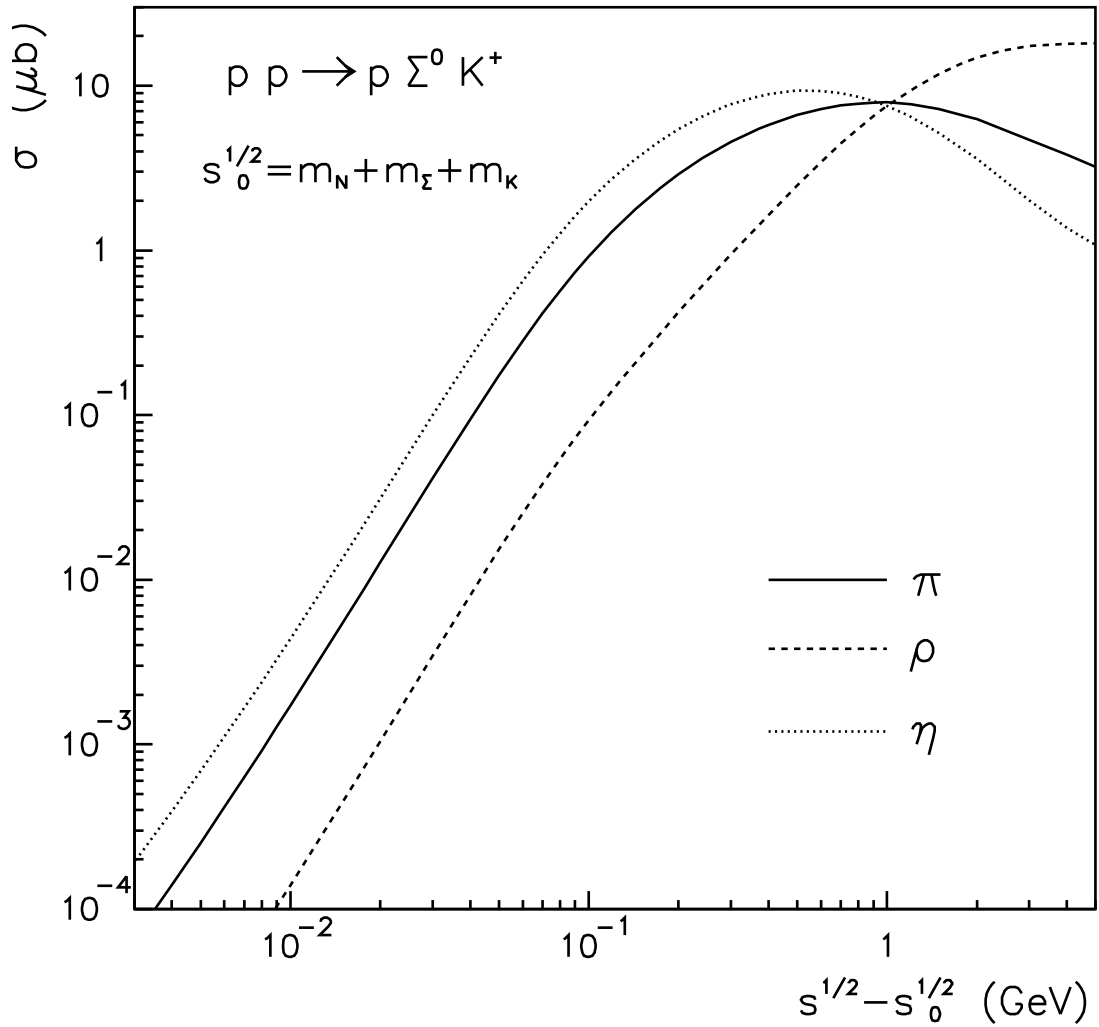,width=16cm}
\caption{\label{kt16p} Separate contributions from $\pi$, $\rho$
and $\eta$-exchanges to the total cross sections  
for the $p p \to p \Sigma^0  K^+$ reaction.}
\end{figure}

Next, we will discuss the $p p \to N \Sigma K$ reactions.
In Fig.~\ref{kt16p} we show the separate contributions from
$\pi$, $\rho$ and $\eta$-exchange to the total cross sections for  
the $p p \to p \Sigma^0 K^+$ reaction.  
The $\rho$-exchange is again dominant at higher energies, while 
the $\eta$ and $\pi$-exchanges are dominant near the threshold. 
The larger contribution from the $\eta$-exchange compared to that of  
$\pi$-exchange, which contrasts with the $p p \to p \Lambda K^+$ 
reaction, is due to the large $\eta N N(1710)$ coupling constant, 
i.e., the large branching ratio of the $N(1710)$ resonance to the
$\eta N$ channel. (See also Tables~\ref{resonance} and~\ref{cconst}.)

\begin{figure}[h]
\epsfig{figure=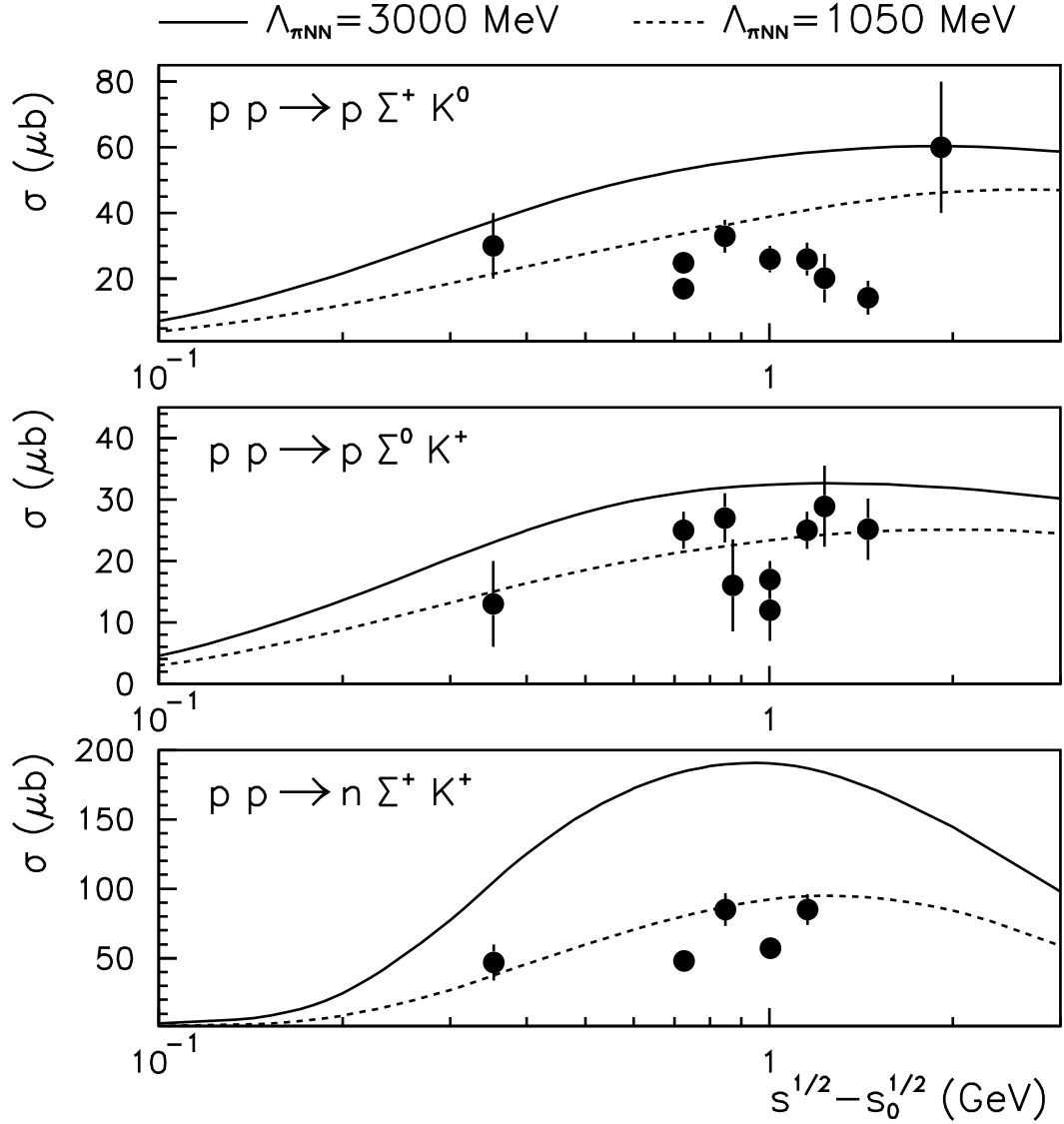,width=16cm}
\caption{\label{kt2p} Energy dependence of the total cross 
sections for the 
$p p \to p \Sigma^+ K^0$,
$p p \to p \Sigma^0 K^+$ and
$p p \to n \Sigma^+ K^+$ reactions.
The dots stand for the experimental 
data from Ref.~\protect\cite{LB} with error bars.
The solid and dashed lines show our results calculated 
using different values for the cut-off parameter, 
$\Lambda_{\pi N N} = 3000$ MeV and $1050$ MeV, respectively.}
\end{figure}

The energy dependence of the total cross sections for the
$p p \to p \Sigma^+ K^0$, $p p \to p \Sigma^0 K^+$ and 
$p p \to n \Sigma^+ K^+$ reactions is shown in Fig.~\ref{kt2p}, 
together with the experimental data~\cite{LB}.
The threshold is, $\sqrt{s_0} = m_N + m_\Sigma+m_K$, 
with $m_\Sigma$ the mass of the $\Sigma$-hyperon.
The solid and dashed lines indicate the results calculated using the two 
different values the cut-off parameter, $\Lambda_{\pi N N} = 3000$ MeV 
and $1050$ MeV, respectively. Although the deviations of the experimental
data are relatively large, our model with the adopted cut-off parameter 
value, $\Lambda_{\pi N N} = 1050$ MeV, reproduces the data fairly well 
using the values, $g^2_{\rho N N}/4 \pi = 0.74$ and $\Lambda_\rho = 920$ 
MeV, fixed in the $p p \to p \Lambda K^+$ reaction.

\begin{figure}[hbt]
\epsfig{figure=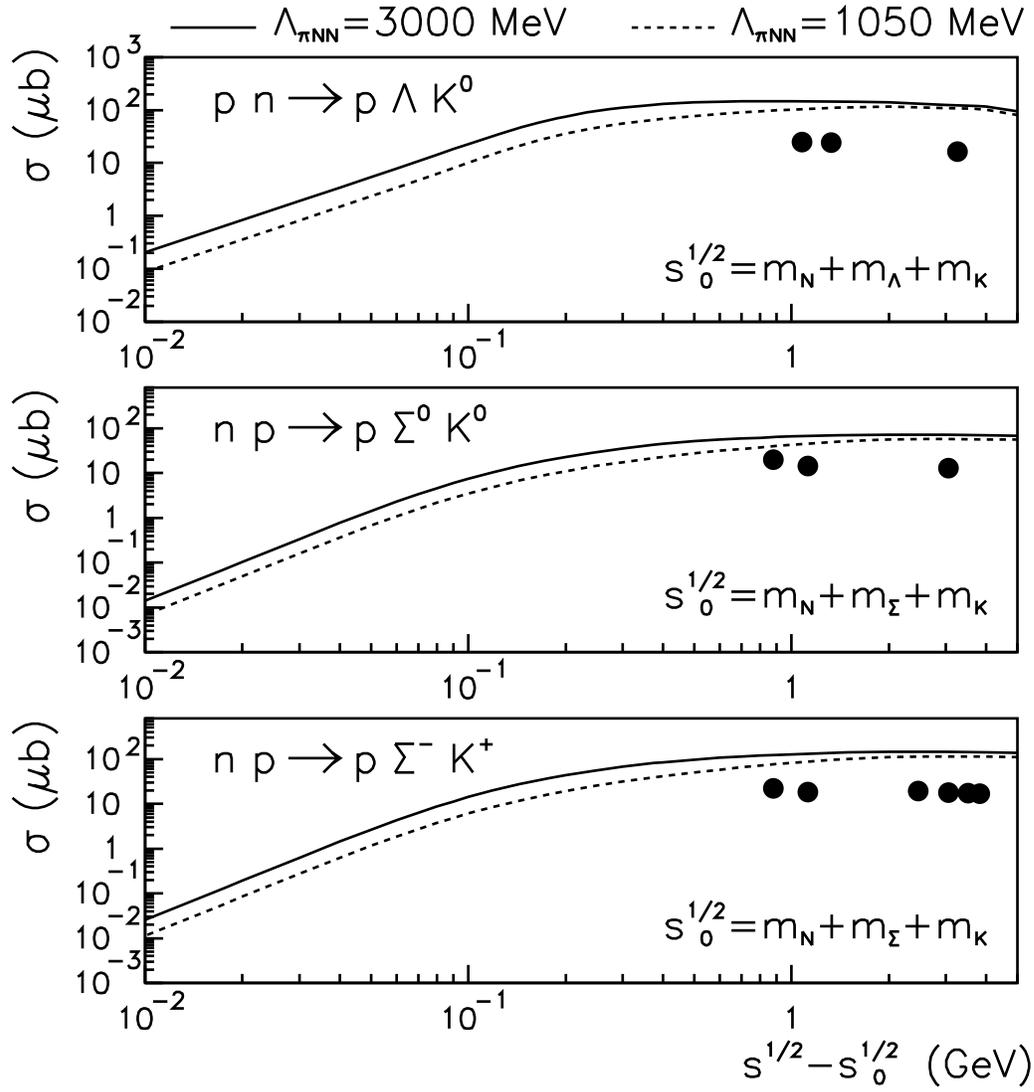,width=16cm}
\caption{\label{kt8p} Same as in Fig.~\protect\ref{kt2p}, but for the 
$pn \to p \Lambda  K^0$, $np \to p \Sigma^0 K^0$ and  
$np \to p \Sigma^- K^+$ reactions. Error bars are smaler than the dots.}
\end{figure}

In Fig.~\ref{kt8p}, we show energy dependence of the    
total cross sections for the $p n \to p \Lambda K^0$, 
$n p \to p \Sigma^0 K^0$ and $n p \to p \Sigma^- K^+$ 
reactions together with the experimental data~\cite{LB}.
The results are again shown for the two values of $\Lambda_{\pi N N}$, 
where the results calculated with the value, $\Lambda_{\pi N N} = 1050$ MeV 
(the dashed lines in Fig.~\ref{kt8p}) should be compared with the data.
At energies about 1 GeV above the threshold, 
the magnitudes of the total cross sections calculated using 
the two different values of the cut-off, 
$\Lambda_{\pi N N}$, become almost equal. This is because, at these  
energies, the momentum of the exchanged pion becomes large and 
the $\pi N N$ form factor~\cite{mac} is insensitive to the value 
of the cut-off parameter, $\Lambda_{\pi N N}$.
The model results (the dashed lines) overestimate the experimental 
data roughly a factor 2, or at most a factor of 7 for one data point, 
although the same parameter set reproduces the data for the
$p p \to p \Lambda K^+$ reaction fairly well.
However, the overestimate of about a factor 2
for the $p n \to p \Lambda K^0$ reaction, can be contrasted 
with the finding of F\"aldt and Wilkin~\cite{wilkin},
that the results for the reaction with a neutron target
differs by factors of 5 to 10. Their calculation included  
the $N(1650)$ resonance alone with pion exchange, 
which is included as one of the contributions in the present 
calculation. 

We should comment about the discrepancies between the calculated 
results and the neutron data.
Our calculations for the $p n (n p) \to N Y K$ reactions are based 
on SU(2) symmetry which should be very good for the present purpose.
For the given Lagrangian densities, which are explained in Sec. 2, 
the calculations for the $p n (n p) \to N Y K$ reactions are 
trivially related to the $p p \to N Y K$ reactions through isospin 
symmetry. We do not have any new mechanism to introduce in the model 
in calculating the $p n (n p) \to N Y K$ reactions -- except 
for charge symmetry breaking, which cannot be expected  
to lead to such a huge difference between the 
$p n (n p) \to N Y K$ and $p p \to N Y K$ reactions.
We should also comment that the data for the $p n (n p) \to N Y K$ 
reactions tabulated in Ref.~\cite{LB} are taken from Ref.~\cite{ans}, 
whose data contain a systematic uncertainty of $\sim 13$ \% in 
the neutron beam normalization, where that error is not  
included in Fig.~\ref{kt8p}.  
Furthermore, their analysis~\cite{ans} was 
based on a number of hypotheses, which in principle, could be 
strongly model dependent.
Thus, we would like to emphasize that it is necessary to investigate further  
the reactions involving a neutron in the initial state,  
both theoretically and experimentally.

The total cross sections for the $K^0$ production channels  
are obtained by
\begin{eqnarray}
\sigma (p p \to p \Lambda K^+)
&=& \sigma (n n \to n \Lambda K^0),\\
\sigma (p n \to n \Lambda K^+)
&=& \sigma (n p \to p \Lambda K^0),
\end{eqnarray}
\begin{eqnarray}
\sigma (p p \to p \Sigma^0 K^+)
&=& \sigma (n n \to n \Sigma^0 K^0),\\
\sigma (p p \to n \Sigma^+ K^+)
&=& \sigma (n n \to p \Sigma^- K^0),\\
\sigma (p n \to n \Sigma^0 K^+)
&=& \sigma (n p \to p \Sigma^0 K^0),\\
\sigma (n p \to p \Sigma^- K^+)
&=& \sigma (p n \to n \Sigma^+ K^0),\\
\sigma (n n \to n \Sigma^- K^+)
&=& \sigma (p p \to p \Sigma^+ K^0).
\end{eqnarray}
%

%%%%%%%%%%%%%%%%%%%%%%%%%%%%%%%%%%%%%%%%%%%%%%%%%%%%%%%%%%%%%%%%%%%%%%%%%%%%%
\subsection{$N N \to \Delta Y K$}

One expects that contributions from the 
$N N \to \Delta Y K$ reactions to the kaon yield in heavy ion collisions 
are small, because of the high threshold energy.
Thus, usually these reactions are not included 
in simulation codes. However, there has been no explicit 
theoretical estimate for the reactions.

\begin{figure}[hbt]
\epsfig{figure=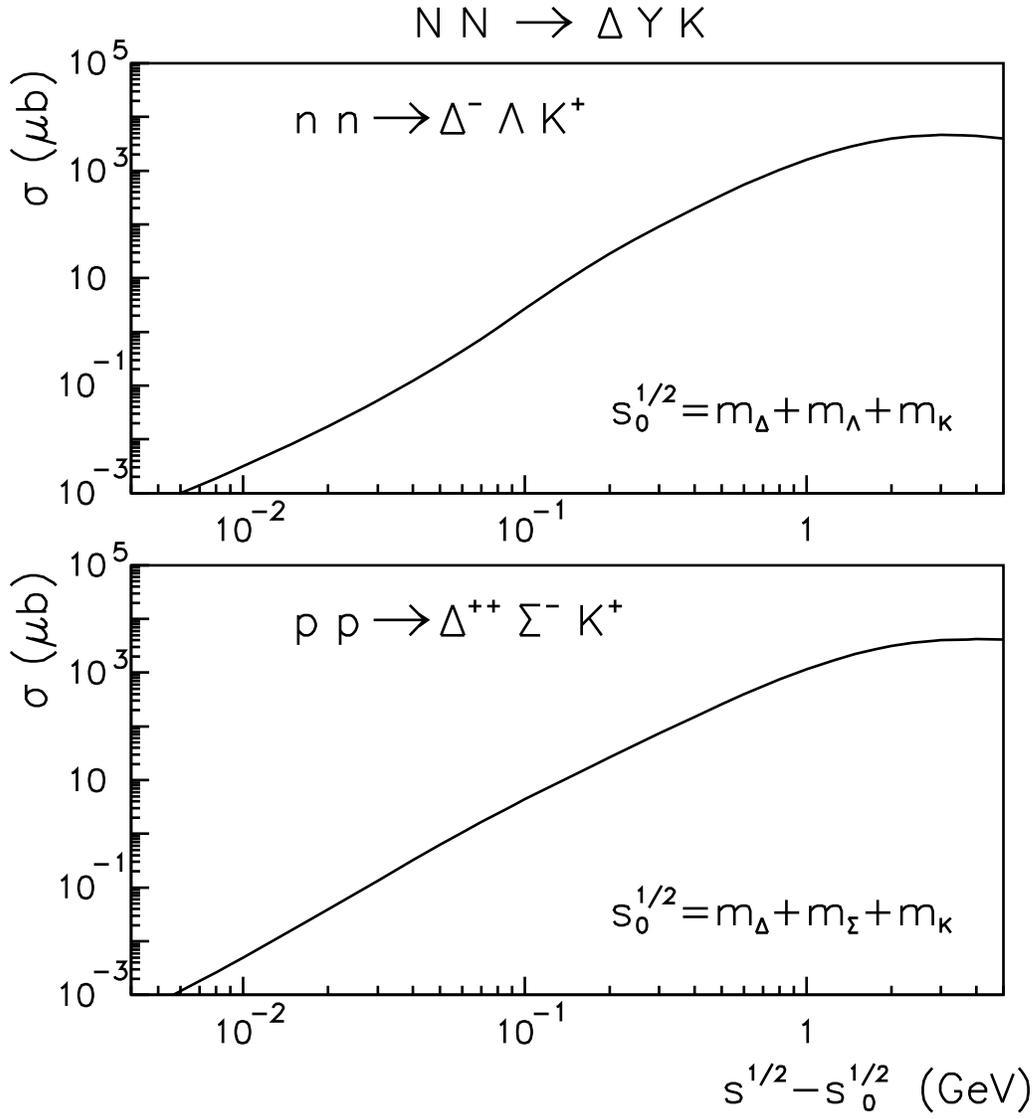,width=16cm}
\caption{\label{kt14p} Energy dependence of the total cross sections 
for the $n n \to \Delta^- \Lambda K^+$ and 
$pp \to \Delta^{++} \Sigma^- K^+$ reactions.}
\end{figure}

In Fig.~\ref{kt14p} we show the   
energy dependence of the total cross sections for the 
$n n \to \Delta^- \Lambda K^+$ and
$p p \to \Delta^{++} \Sigma^- K^+$ reactions.
At $N N$ invariant collision energies about 3.1 GeV, the magnitudes of  
the total cross sections for both reactions become about $70 \mu b$.  
These magnitudes are comparable with that of the $p p \to p \Lambda K^+$ 
reaction at these energies. However, they become about 10 times larger 
than those of the $p p \to p \Lambda K^+$ reaction at $N N$ invariant 
collision energies about 3.5 GeV.
Thus, these reactions, which are usually discarded in theoretical 
studies of kaon production in heavy ion collisions, 
might give large contributions to the kaon yield, if there 
are plenty of $N N$ pairs which have such  
collision energies.

\begin{figure}[hbt]
\epsfig{figure=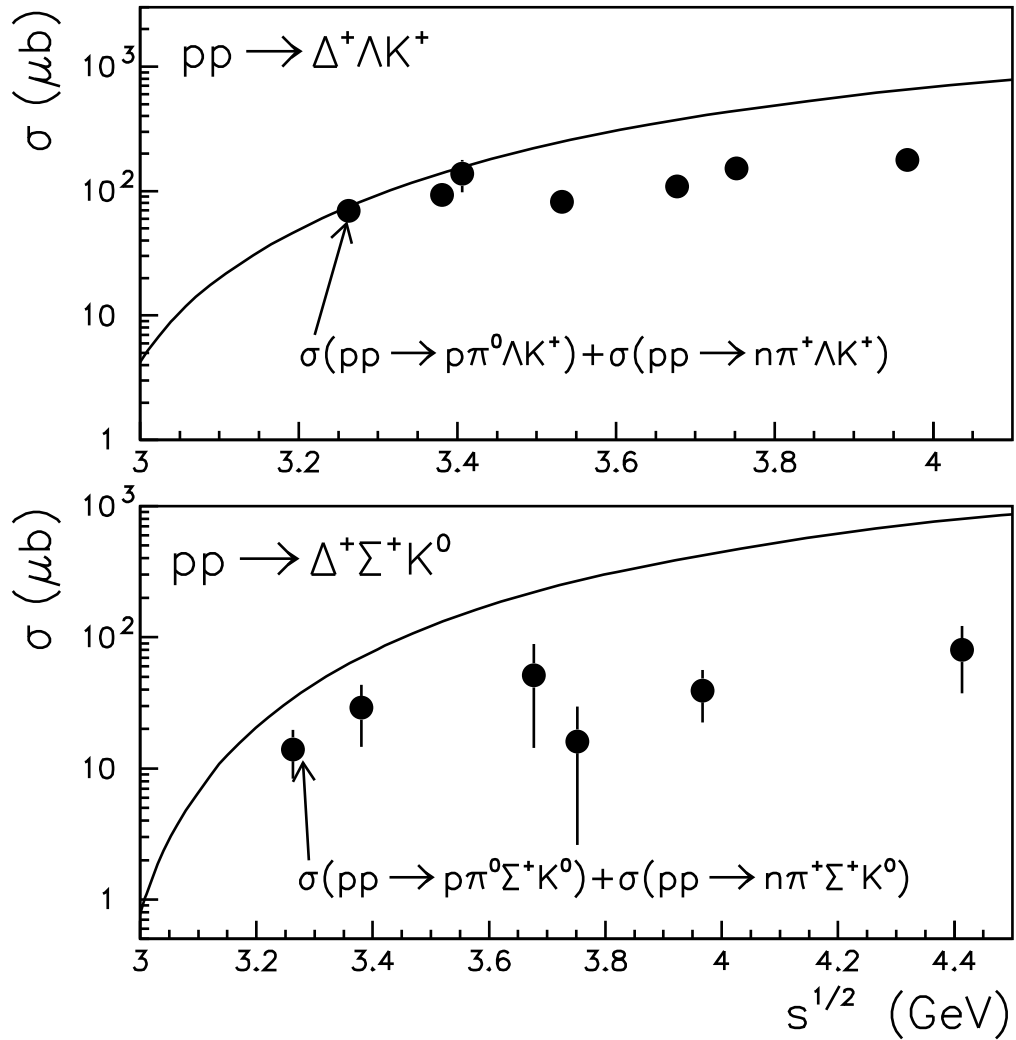,width=16cm}
\caption{\label{kt14pt} Comparison of the energy dependence of  
the total cross sections for the $pp \to \Delta Y K$ and 
$pp \to N \pi Y K$ reactions. Note that the parametrizations 
are recommended to be used up to invariant collision energy about 
3.6 GeV.}
\end{figure}

In order to make an estimate of the uncertainty in the calculation  
for these reactions, we make a comparison with the data for the 
$pp \to N \pi Y K$ reactions in Fig.~\ref{kt14pt}. 
Note that these reactions have different thresholds, but the comparison 
should be helpful in judging whether the results  
contain more than a order of magnitude uncertainties.
We restrict ourselves to $NN$ invariant collision energies below 
about 3.5 GeV (as we do throughout this paper) the predictions are 
certainly of the same order of magnitude as the data and therefore 
not unreasonable. We conclude that 
the $NN \to \Delta Y K$ reactions may be equally as 
important as the $NN \to NYK$ reactions for heavy ion simulations.
However, this argument should be checked eventually by heavy ion simulations, 
keeping in mind the theoretical uncertainties.

The total cross sections for the other isospin channels are obtained by
\begin{eqnarray}
\sigma(nn \to \Delta^- \Lambda K^+)
&=& \sigma(pp \to \Delta^{++} \Lambda K^0) \nn \\
= 3 \sigma(pn \to \Delta^0 \Lambda K^+) 
&=& 3 \sigma(np \to \Delta^+ \Lambda K^0) \nn \\
= 3 \sigma(pp \to \Delta^+ \Lambda K^+) 
&=& 3 \sigma(nn \to \Delta^0 \Lambda K^0), 
\end{eqnarray}
\begin{eqnarray}
\sigma(pp\to \Delta^{++} \Sigma^- K^+) 
&=& \sigma(nn\to \Delta^- \Sigma^+ K^0) \nn \\
= 2\sigma(nn\to \Delta^- \Sigma^0 K^+) 
&=& 2\sigma(pp\to \Delta^{++} \Sigma^0 K^0) \nn \\
= 3\sigma(np\to \Delta^+ \Sigma^- K^+) 
&=& 3\sigma(pp\to \Delta^+ \Sigma^+ K^0) \nn \\
= 3\sigma(nn\to \Delta^0 \Sigma^- K^+)
&=& 3\sigma(pn\to \Delta^0 \Sigma^+ K^0) \nn \\
= 6\sigma(pp\to \Delta^+ \Sigma^0 K^+) 
&=& 6\sigma(np\to \Delta^+ \Sigma^0 K^0) \nn \\
= 6\sigma(pn\to \Delta^0 \Sigma^0 K^+) 
&=& 6\sigma(nn\to \Delta^0 \Sigma^0 K^0).
\end{eqnarray}

%%%%%%%%%%%%%%%%%%%%%%%%%%%%%%%%%%%%%%%%%%%%%%%%%%%%%%%%%%%%%%%%%%%%%%%%%%%%%
\subsection{$\Delta N \to N Y K$}

According to the simulation results~\cite{lan,hua2,har2,bali,libbk}, 
it is widely believed that kaons are mostly produced through the 
$\Delta N \to N Y K$ and $\Delta \Delta \to N Y K$ reactions 
in heavy ion collisions. 
However, in most of the theoretical studies with 
the Boltzmann/Vlasov-Uehling-Uhlenbeck approach 
(BUU/VUU)~\cite{cas1,buu,zwe,cas3,bar,fan1}, 
or quantum molecular dynamics (QMD)~\cite{rqmd,qmd}, the 
parametrizations used for these total cross sections 
have been made in a simple phenomenological manner in free 
space~\cite{schu,ran,cug1}. In addition, 
the parametrizations for all isospin channels are obtained 
by fitting to the existing limited isospin channels of the data. 
Furthermore, those parametrizations for the reactions involving a $\Delta$ 
or $\Delta$'s used to draw the conclusion~\cite{lan,hua2,har2,bali,kaos,rus}, 
are obtained using the cross sections for the nucleon except for the isospin 
difference~\cite{ran}.
In addition, the coupling constants relevant for 
the $\Delta$ are assumed to be equal to those for the nucleon, 
e.g., $F_{N N \pi} = F_{N \Delta \pi}$ and 
$F_{\Delta \Delta \pi} = F_{N N \pi}$. 

Recently, this isospin symmetry ansatz 
suggested in Ref.~\cite{ran} has been reconsidered Li and Ko~\cite{libbk},  
and elementary kaon production cross sections in baryon baryon 
reactions have been calculated using a one pion and one kaon exchange model.
But their approach is still not fully consistent 
because they adopted the results of the pion baryon reactions from 
the resonance model~\cite{tsu}, where the resonance model~\cite{tsu} does
not include one kaon exchange in its basic assumptions.
Thus, if one wants to be consistent with the parameters determined in 
the pion baryon reactions, one kaon exchange should not be introduced.  

\begin{figure}[hbt]
\epsfig{figure=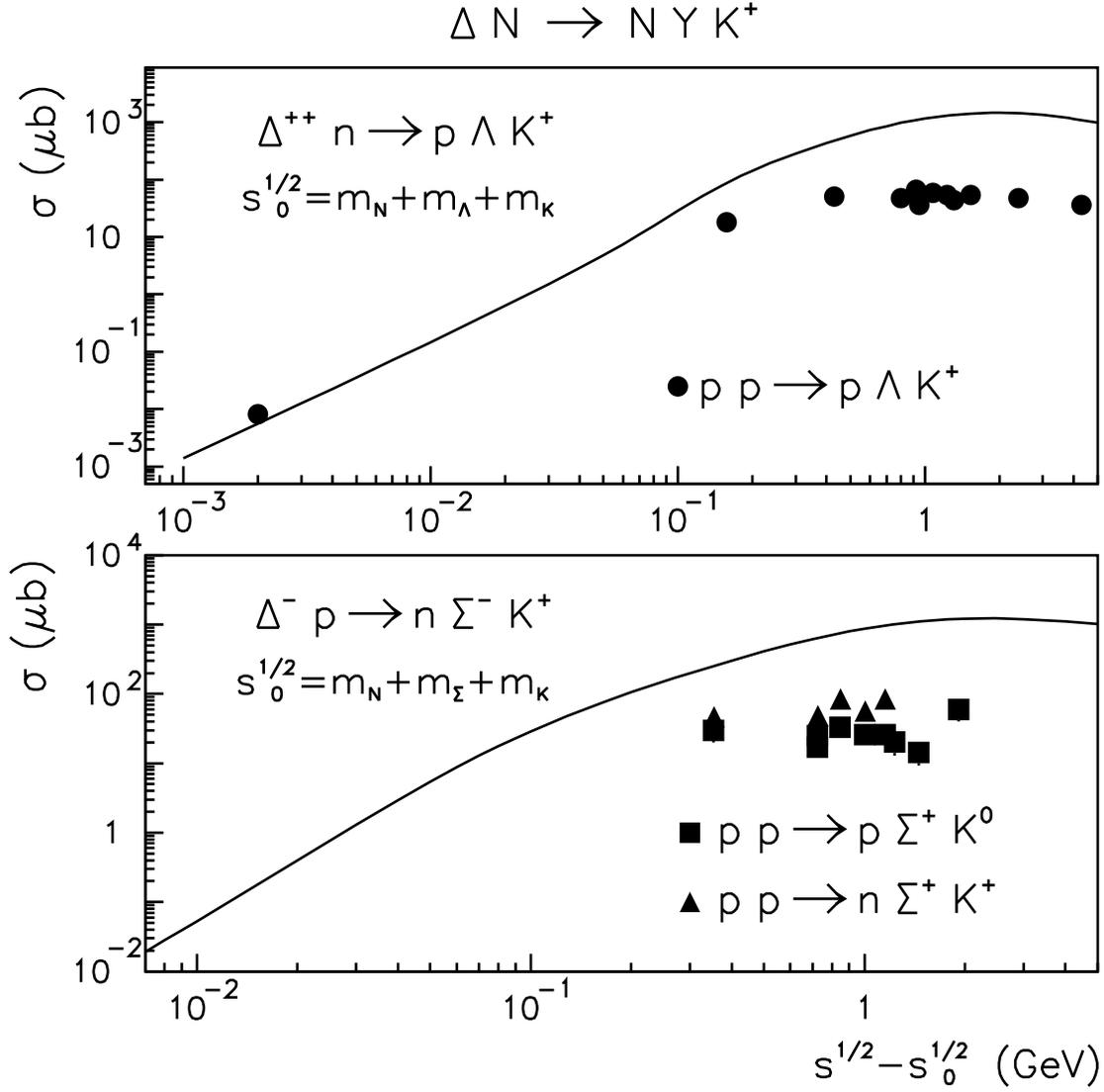,width=16cm}
\caption{\label{kt13p} Energy dependence of the total 
cross sections for the $\Delta^{++} n \to p \Lambda K^+$ and 
$\Delta^- p \to n \Sigma^- K^+$ reactions. The dots, triangles 
and squares are the experimental data  
for the $p p \to p \Lambda K^+$,
$p p \to p \Sigma^+ K^0$ and $p p \to n \Sigma^+ K^+$ 
reactions, respectively, from Ref.~\protect\cite{LB}.}
\end{figure}

In Fig.~\ref{kt13p} we show the energy dependence of the total cross sections 
for the $\Delta^{++} n \to p \Lambda K^+$ and 
$\Delta^- p \to n \Sigma^- K^+$ reactions, together with the data 
for the $p p \to p \Lambda K^+$, $p p \to p \Sigma^+ K^0$
and $p p \to n \Sigma^+ K^+$ reactions~\cite{LB} plotted at 
the same excess energy above each threshold energy corresponding 
to the $\Lambda$ and $\Sigma$ production channels. 
The magnitude of the total cross section for the
$\Delta^{++} n \to p \Lambda K^+$ reaction is almost equal to 
that of the $p p \to p \Lambda K^+$ reaction at the same excess 
energy just above the threshold, $\sqrt{s}-\sqrt{s_0} < 100$~MeV. 
However, at energies $\sqrt{s}-\sqrt{s_0} \simeq 1$ GeV, 
the magnitudes of both the $\Delta^{++} n \to p \Lambda K^+$ 
and $\Delta^- p \to n \Sigma^- K^+$ reactions become 
larger by about a factor of 6 than those of 
the $p p \to p \Lambda K^+$, and $p p \to p \Sigma^+ K^0$ and 
$p p \to n \Sigma^+ K^+$ reactions, respectively. 
One of the main reasons for this is the different spin 
structure of nucleon and $\Delta$, 
as can be seen from Eq. (\ref{pmunu}), 
which produces different energy dependence in the Lorentz invariant 
scattering amplitude. (The difference in form factors and coupling 
constants is also the other reason.)
Our result may be consistent  
with the conclusion that the $\Delta N \to N Y K$ reactions give 
the largest contribution to kaon yield in heavy ion collisions.
However, this should be explicitly checked. 

The total cross sections for the other isospin channels of 
the $\Delta N \to N Y K$ reactions are obtained by:
\begin{eqnarray}
\sigma(\Delta^{++} n \to p \Lambda K^+) 
&=& \sigma(\Delta^- p \to n \Lambda K^0) \nn \\ 
= 3\sigma(\Delta^+ p \to p \Lambda K^+) 
&=& 3\sigma(\Delta^+ n \to p \Lambda K^0) \nn \\
= 3\sigma(\Delta^+ n \to n \Lambda K^+) 
&=& 3\sigma(\Delta^0 p \to p \Lambda K^0) \nn \\  
= 3\sigma(\Delta^0 p \to n \Lambda K^+) 
&=& 3\sigma(\Delta^0 n \to n \Lambda K^0), \\ 
\nn \\
\sigma(\Delta^- p \to n \Sigma^- K^+)
&=& \sigma(\Delta^{++} n \to p \Sigma^+ K^0) \nn \\
= 2\sigma(\Delta^{++} n \to p \Sigma^0 K^+) 
&=& 2\sigma(\Delta^- p \to n \Sigma^0 K^0) \nn \\
= 3\sigma(\Delta^+ n \to p \Sigma^- K^+) 
&=& 3\sigma(\Delta^+ p \to p \Sigma^+ K^0) \nn \\
 = 3\sigma(\Delta^0 p \to p \Sigma^- K^+) 
&=& 3\sigma(\Delta^+ n \to n \Sigma^+ K^0) \nn \\
 = 3\sigma(\Delta^0 n \to n \Sigma^- K^+) 
&=& 3\sigma(\Delta^0 p \to n \Sigma^+ K^0) \nn \\
 = 6\sigma(\Delta^+ p \to p \Sigma^0 K^+) 
&=& 6\sigma(\Delta^+ n \to p \Sigma^0 K^0) \nn \\
 = 6\sigma(\Delta^+ n \to n \Sigma^0 K^+) 
&=& 6\sigma(\Delta^0 p \to p \Sigma^0 K^0) \nn \\
 = 6\sigma(\Delta^0 p \to n \Sigma^0 K^+) 
&=& 6\sigma(\Delta^0 n \to n \Sigma^0 K^0). 
\end{eqnarray}
%

%%%%%%%%%%%%%%%%%%%%%%%%%%%%%%%%%%%%%%%%%%%%%%%%%%%%%%%%%%%%%%%%%%
\subsection{$ \Delta N \to \Delta Y K$}

\begin{figure}[hbt]
\epsfig{figure=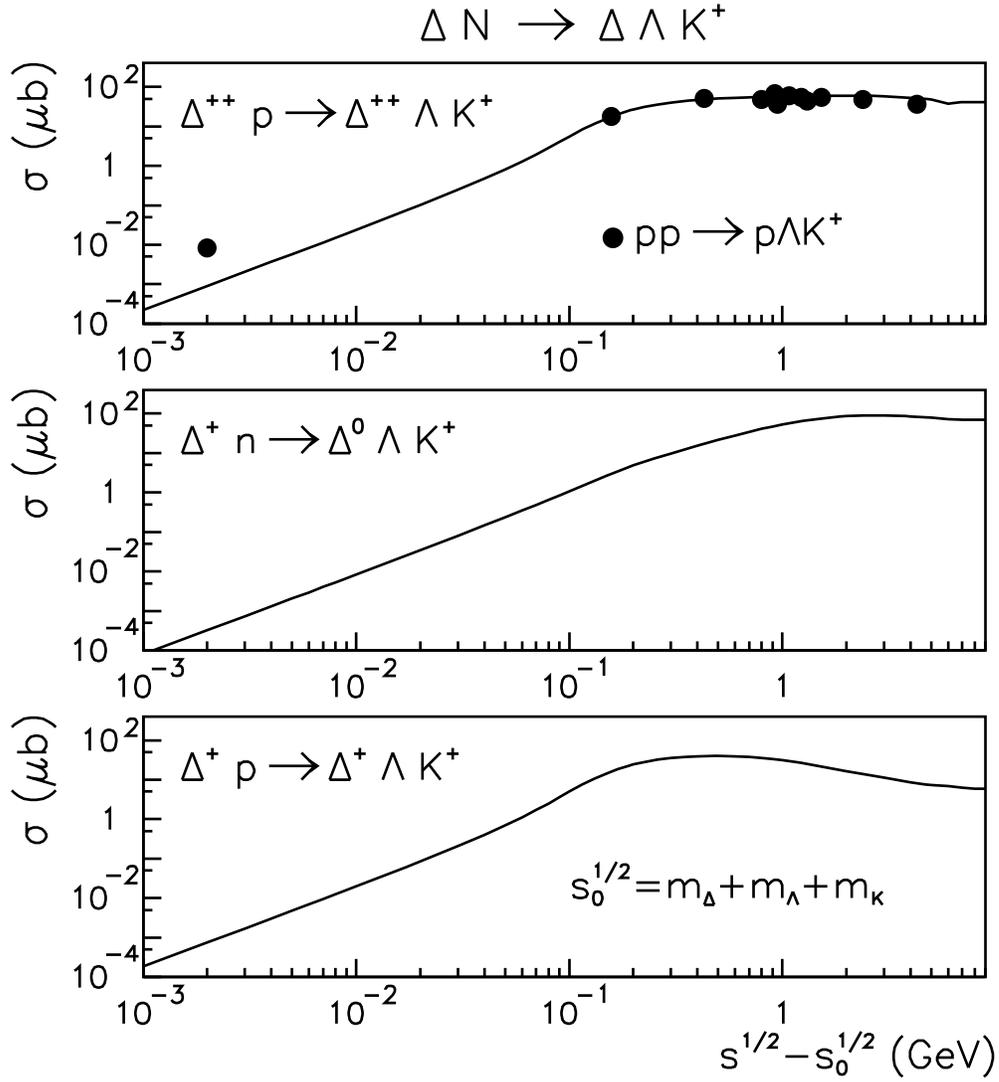,width=16cm}
\caption{\label{kt9p} Energy dependence of the total cross sections 
for the $\Delta^{++} p \to \Delta^{++} \Lambda K^+$, 
$\Delta^+ n \to \Delta^0 \Lambda K^+$ and
$\Delta^+ p \to \Delta^+ \Lambda K^+$ reactions. The dots are the 
experimental data from Ref.~\protect\cite{LB} plotted at the same 
excess energies from each threshold energy.}
\end{figure}

\begin{figure}[hbt]
\epsfig{figure=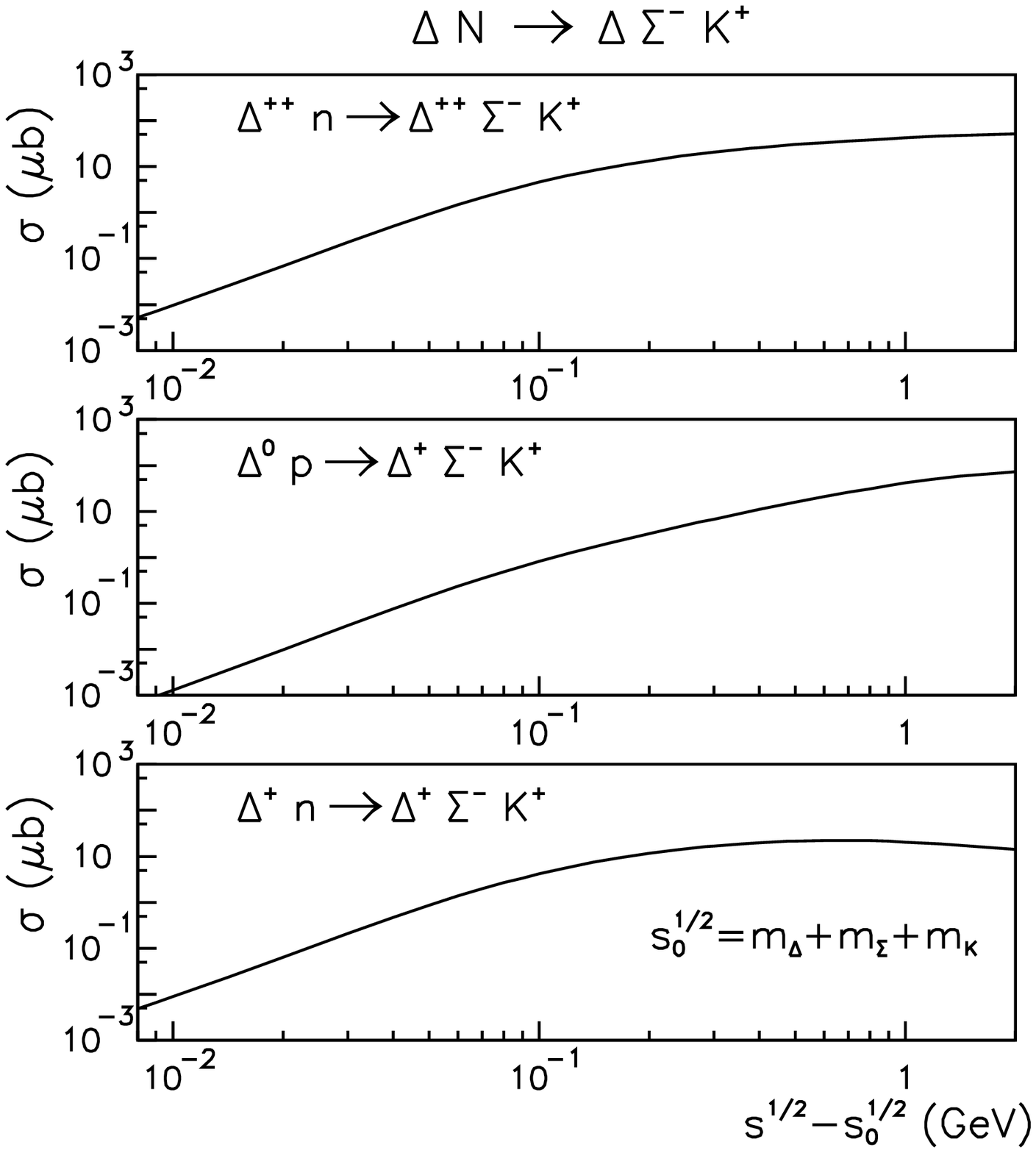,width=16cm}
\caption{\label{kt10p} Energy dependence of the total cross sections 
for the $\Delta^{++} n \to \Delta^{++} \Sigma^- K^+$, 
$\Delta^0 p \to \Delta^+ \Sigma^- K^+$ and
$\Delta^+ n \to \Delta^+ \Sigma^- K^+$ reactions.}
\end{figure}

\begin{figure}[hbt]
\epsfig{figure=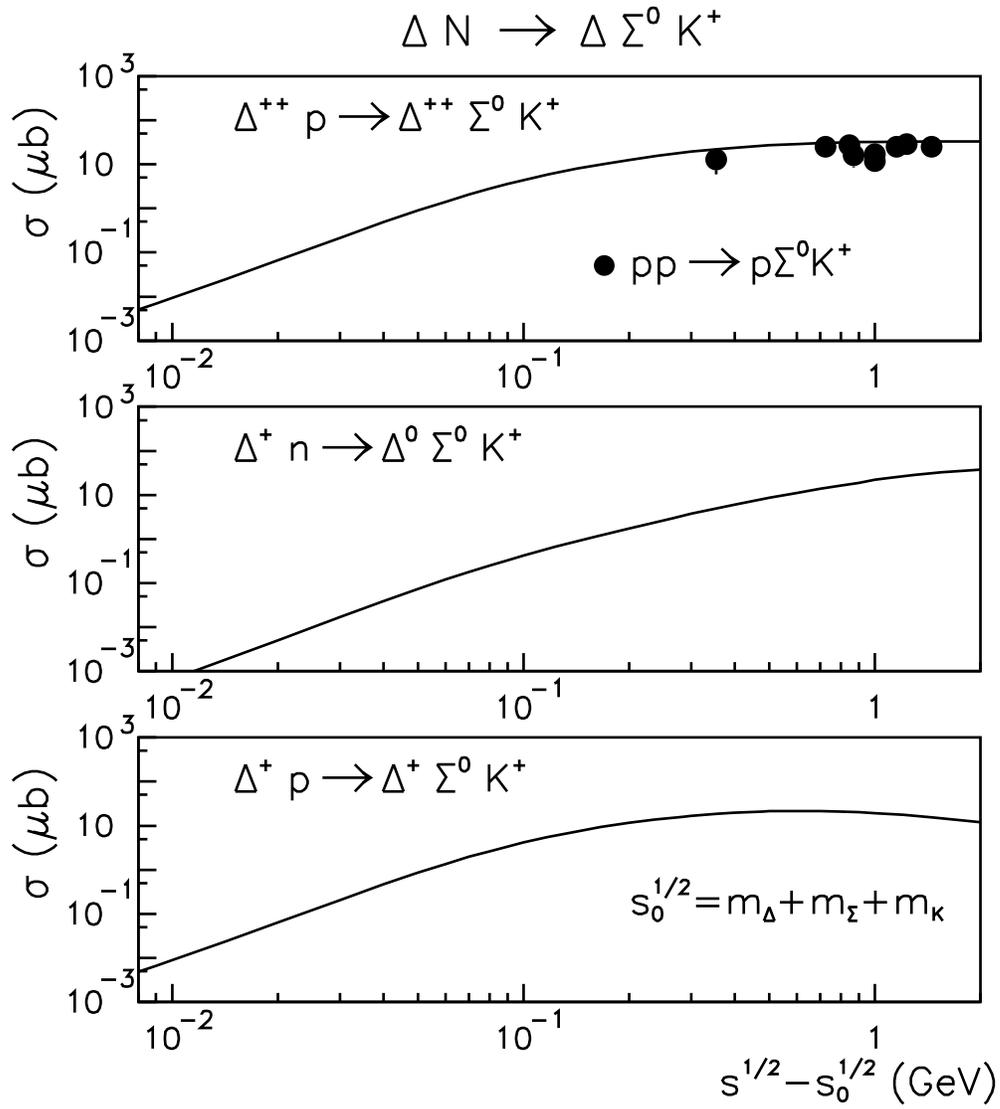,width=16cm}
\caption{\label{kt11p} Same as in Fig.~\protect\ref{kt9p}, but 
for the $\Delta^{++} p \to \Delta^{++} \Sigma^0 K^+$,
$\Delta^+ n \to \Delta^0 \Sigma^0 K^+$ and
$\Delta^+ p \to \Delta^+ \Sigma^0 K^+$ reactions, and the data 
for the $p p \to p \Sigma^0 K^+$ reaction.}
\end{figure}

\begin{figure}[hbt]
\epsfig{figure=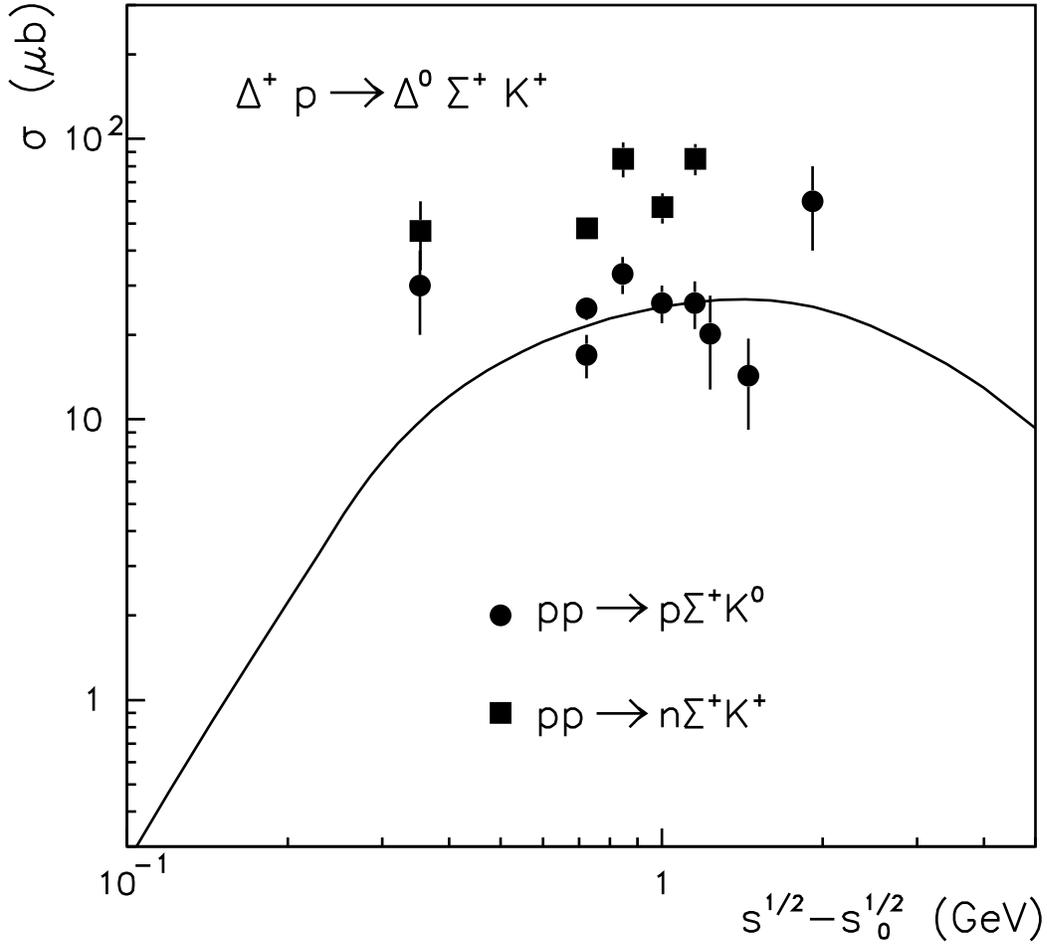,width=16cm}
\caption{\label{kt12p} Energy dependence of the total cross sections 
for the $\Delta^+ p \to \Delta^0 \Sigma^+ K^+$ reaction 
together with the experimental data from Ref.~\protect\cite{LB} 
for the $p p \to p \Sigma^+ K^0$ and $p p \to n \Sigma^+ K^+$ reactions,  
plotted at the same excess energies.  
The threshold for the $\Delta^+ p \to \Delta^0 \Sigma^+ K^+$
reaction is $\protect\sqrt{s_0}=m_\Delta + m_\Sigma + m_K$.}
\end{figure}

Similarly to the situation for the $N N \to \Delta Y K$ reactions, 
the $\Delta \Delta \to \Delta Y K$ reactions  
are not usually included in the simulation codes   
due to the high threshold energy.
However, these reactions may be responsible for the low energy
part of the kaon spectra.  
In Figs.~\ref{kt9p} --~\ref{kt12p}, we show energy dependence of 
the total cross sections sufficient for obtaining all isospin channels 
in the reactions, together with the experimental data of 
the $p p \to p \Lambda K^+$, $p p \to p \Sigma^0 K^+$, 
$p p \to p \Sigma^+ K^0$ and $p p \to n \Sigma^+ K^+$ reactions
plotted at the same excess energies above the corresponding thresholds. 
The magnitudes of the total cross sections are almost equal 
to those measured for the $p p \to N Y K$ reactions 
at the same excess energy, $\sqrt{s}-\sqrt{s_0} > 100$~MeV, 
with respect to each corresponding threshold, $\sqrt{s_0}$.

The total cross sections for the other isospin channels are obtained by:
\begin{eqnarray}
\sigma(\Delta^{++} p \to \Delta^{++} \Lambda K^+) 
&=&\sigma(\Delta^{++} n \to \Delta^{++} \Lambda K^0) \nn \\
 = \sigma(\Delta^- p \to \Delta^- \Lambda K^+) 
&=&\sigma(\Delta^- n \to \Delta^- \Lambda K^0),  
\\ \nn \\ 
3\sigma(\Delta^+ n \to \Delta^0 \Lambda K^+) 
&=& 3\sigma(\Delta^0 p \to \Delta^+ \Lambda K^0) \nn \\   
 = 4\sigma(\Delta^{++} n \to \Delta^+ \Lambda K^+)
&=& 4\sigma(\Delta^+ p \to \Delta^{++} \Lambda K^0) \nn \\   
 = 4\sigma(\Delta^0 n \to \Delta^- \Lambda K^+)  
&=& 4\sigma(\Delta^- p \to \Delta^0 \Lambda K^0), 
\\ \nn \\ 
\sigma(\Delta^+ p \to \Delta^+ \Lambda K^+) 
&=& \sigma(\Delta^+ n \to \Delta^+ \Lambda K^0) \nn \\
 = \sigma(\Delta^0 p \to \Delta^0 \Lambda K^+) 
&=& \sigma(\Delta^0 n \to \Delta^0 \Lambda K^0), 
\\ \nn \\ 
%%%%%%%%%%%%%%
\sigma(\Delta^{++} n \to \Delta^{++} \Sigma^- K^+)
&=&\sigma(\Delta^{++} p \to \Delta^{++} \Sigma^+ K^0) \nn \\ 
 = \sigma(\Delta^- n \to \Delta^- \Sigma^- K^+)  
&=& \sigma(\Delta^- p \to \Delta^- \Sigma^+ K^0), 
\\ \nn \\ 
3\sigma(\Delta^0 p \to \Delta^+ \Sigma^- K^+) 
&=& 3\sigma(\Delta^+ n \to \Delta^0 \Sigma^+ K^0) \nn \\
 = 4\sigma(\Delta^+ p \to \Delta^{++} \Sigma^- K^+)
&=& 4\sigma(\Delta^{++} n \to \Delta^+ \Sigma^+ K^0) \nn \\
 = 4\sigma(\Delta^- p \to \Delta^0 \Sigma^- K^+) 
&=& 4\sigma(\Delta^0 n \to \Delta^- \Sigma^+ K^0),  
\\ \nn \\ 
\sigma(\Delta^+ n \to \Delta^+ \Sigma^- K^+) 
&=& \sigma(\Delta^+ p \to \Delta^+ \Sigma^+ K^0) \nn \\
 = \sigma(\Delta^0 n \to \Delta^0 \Sigma^- K^+) 
&=& \sigma(\Delta^0 p \to \Delta^0 \Sigma^+ K^0), 
\\ \nn \\ 
\sigma(\Delta^{++} p \to \Delta^{++} \Sigma^0 K^+) 
&=& \sigma(\Delta^{++} n \to \Delta^{++} \Sigma^0 K^0) \nn \\
 = \sigma(\Delta^- p \to \Delta^- \Sigma^0 K^+) 
&=& \sigma(\Delta^- n \to \Delta^- \Sigma^0 K^0), 
\\ \nn \\ 
3\sigma(\Delta^+ n \to \Delta^0 \Sigma^0 K^+) 
&=& 3\sigma(\Delta^0 p \to \Delta^+ \Sigma^0 K^0) \nn \\
 = 4\sigma(\Delta^{++} n \to \Delta^+ \Sigma^0 K^+)
&=& 4\sigma(\Delta^+ p \to \Delta^{++} \Sigma^0 K^0) \nn \\
 = 4\sigma(\Delta^0 n \to \Delta^- \Sigma^0 K^+) 
&=& 4\sigma(\Delta^- p \to \Delta^0 \Sigma^0 K^0), 
\\ \nn \\ 
\sigma(\Delta^+ p \to \Delta^+ \Sigma^0 K^+) 
&=& \sigma(\Delta^+ n \to \Delta^+ \Sigma^0 K^0) \nn \\ 
 = \sigma(\Delta^0 p \to \Delta^0 \Sigma^0 K^+) 
&=& \sigma(\Delta^0 n \to \Delta^0 \Sigma^0 K^0),
\\ \nn \\ 
3\sigma(\Delta^+ p \to \Delta^0 \Sigma^+ K^+) 
&=& 3\sigma(\Delta^0 n \to \Delta^+ \Sigma^- K^0) \nn \\
 = 4\sigma(\Delta^{++} p \to \Delta^+ \Sigma^+ K^+)
&=& 4\sigma(\Delta^+ n \to \Delta^{++} \Sigma^- K^0) \nn \\
 = 4\sigma(\Delta^0 p \to \Delta^- \Sigma^+ K^+) 
&=& 4\sigma(\Delta^- p \to \Delta^0 \Sigma^0 K^0).
\end{eqnarray}
%

%%%%%%%%%%%%%%%%%%%%%%%%%%%%%%%%%%%%%%%%%%%%%%%%%%%%%%%%%%%%%%%%%%%%%%%%%
\subsection{$ \Delta \Delta \to \Delta Y K$}

\begin{figure}[hbt]
\epsfig{figure=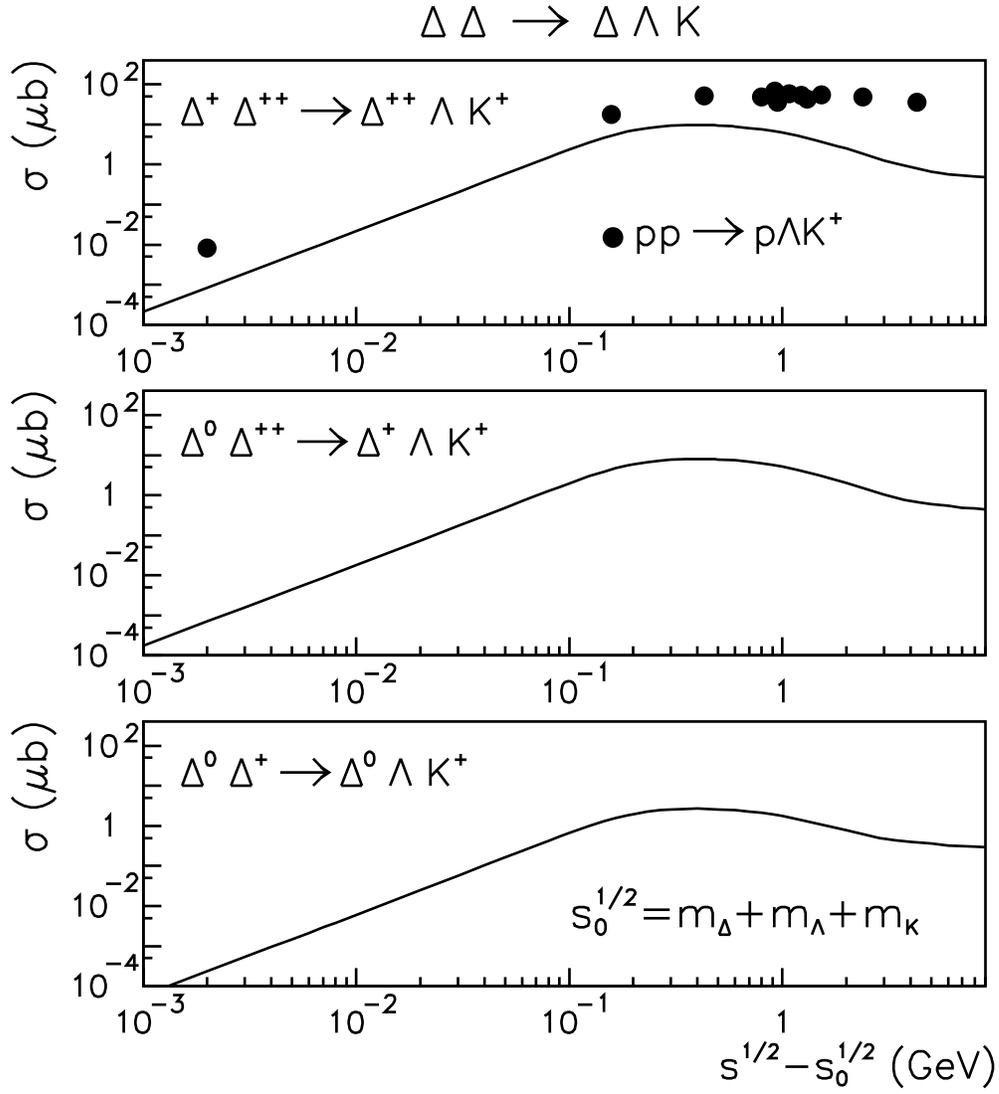,width=16cm}
\caption{\label{kt17p} Same as in Fig.~\protect\ref{kt9p}, but 
for the $\Delta^+ \Delta^{++} \to \Delta^{++} \Lambda K^+$,
$\Delta^0 \Delta^{++} \to \Delta^+ \Lambda K^+$ and
$\Delta^0 \Delta^+ \to \Delta^0 \Lambda K^+$ reactions.}
\end{figure}

\begin{figure}[hbt]
\epsfig{figure=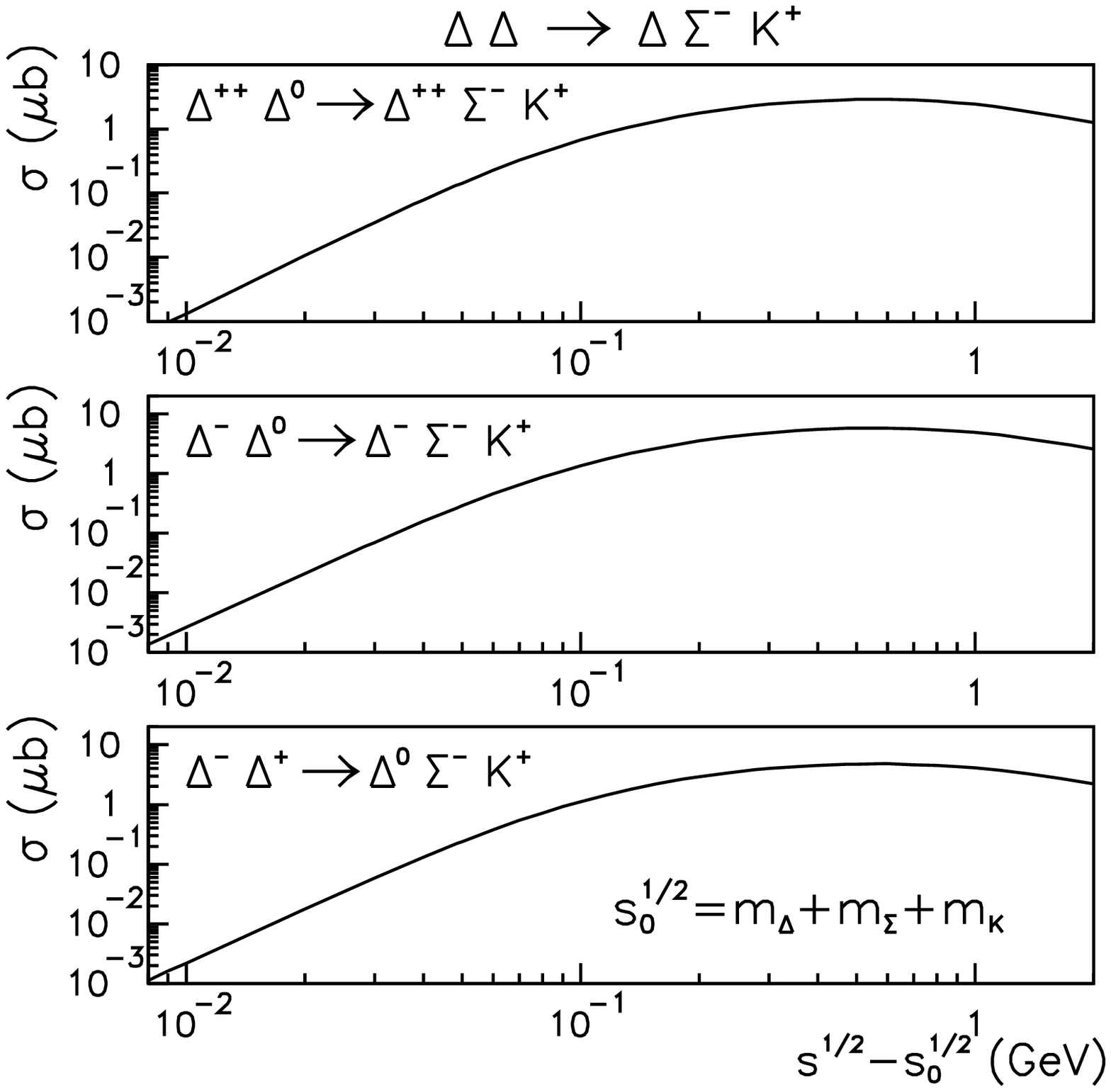,width=16cm}
\caption{\label{kt18p} Energy dependence of the total cross sections 
for the $\Delta^{++} \Delta^0 \to \Delta^{++} \Sigma^- K^+$,
$\Delta^- \Delta^0 \to \Delta^- \Sigma^- K^+$ and
$\Delta^- \Delta^+ \to \Delta^0 \Sigma^- K^+$ reactions.}
\end{figure}

\begin{figure}[hbt]
\epsfig{figure=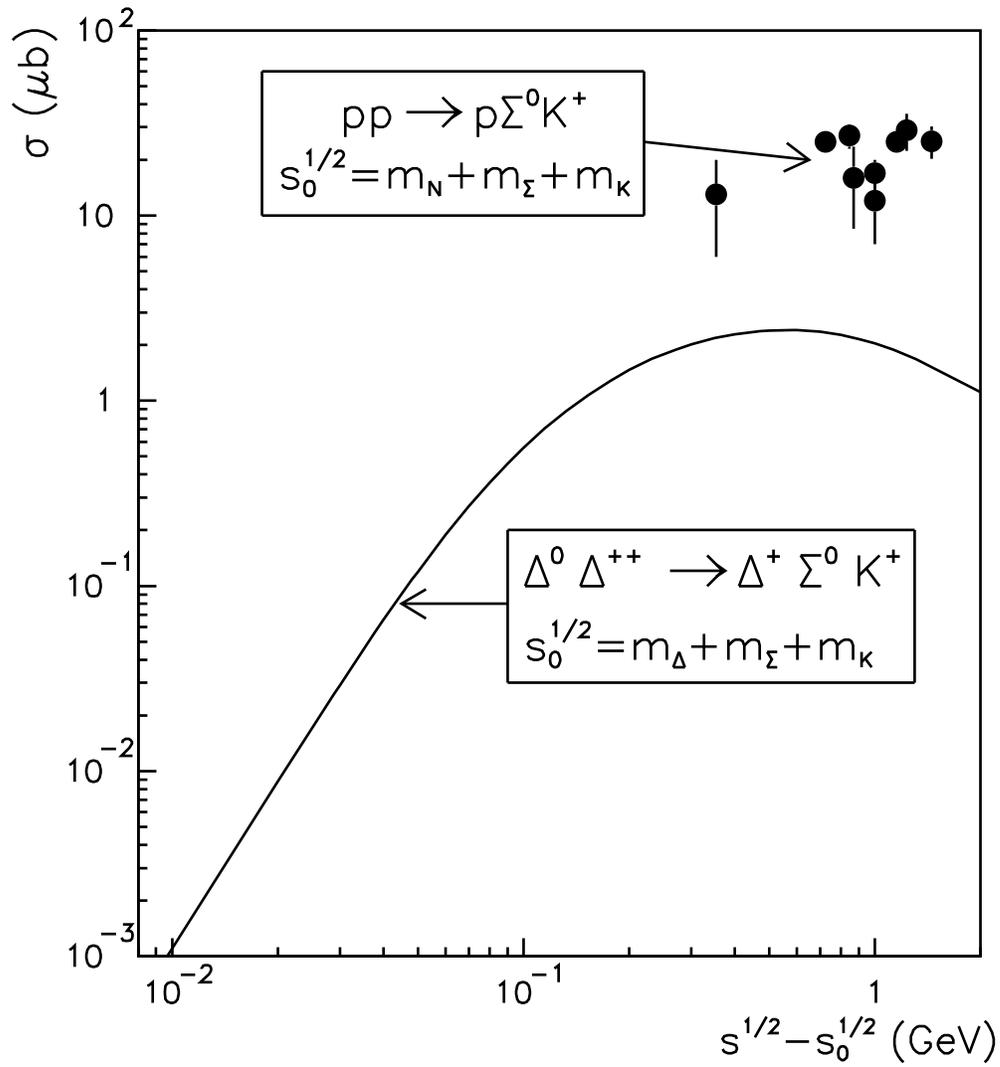,width=16cm}
\caption{\label{kt19p} Same as in Fig.~\protect\ref{kt12p} but 
for the $\Delta^0 \Delta^{++}\to \Delta^+ \Sigma^0 K^+$ reaction and 
the experimental data are for the $p p \to p \Sigma^0 K^+$ reaction.}
\end{figure}

The energy dependence of the total cross sections for the 
$\Delta \Delta \to \Delta \Lambda K$ and
$\Delta \Delta \to \Delta \Sigma K$ reactions are shown in 
Figs.~\ref{kt17p},~\ref{kt18p} and \ref{kt19p}, together with 
the experimental data for the $p p \to p \Lambda K^+$ and 
$p p \to p \Sigma^0 K^+$ reactions plotted at the same excess energies 
above the corresponding hyperon production threshold.
As expected, the magnitudes of the total cross sections for these 
reactions are small, because of the high threshold energies
and three $\Delta$'s are involved in the processes.
The magnitudes for these reactions are almost 10 times smaller than those 
for the $p p \to p \Lambda K^+$ reaction which is shown in 
Fig.~\ref{kt19p}. The contributions from these reactions 
to the kaon yield in heavy ion collisions are therefore negligible. 

The total cross sections for the other isospin channels are obtained by:
\bg
\sigma(\Delta^+ \Delta^{++} \to \Delta^{++} \Lambda K^+) 
&=& \sigma(\Delta^0 \Delta^- \to \Delta^- \Lambda K^0) \nn \\ 
 = 3\sigma(\Delta^0 \Delta^0 \to \Delta^- \Lambda K^+) 
&=& 3\sigma(\Delta^+ \Delta^+ \to \Delta^{++} \Lambda K^0) \nn \\ 
 = 9\sigma(\Delta^+ \Delta^+ \to \Delta^+ \Lambda K^+)
&=& 9\sigma(\Delta^0 \Delta^0 \to \Delta^0 \Lambda K^0), 
\\ \nn \\
\sigma(\Delta^0 \Delta^{++} \to \Delta^+ \Lambda K^+) 
&=& \sigma(\Delta^+ \Delta^- \to \Delta^0 \Lambda K^0), 
\\ \nn \\
\sigma(\Delta^0 \Delta^+ \to \Delta^0 \Lambda K^+)
&=& \sigma(\Delta^+ \Delta^0 \to \Delta^+ \Lambda K^0),
\\ \nn \\
\sigma(\Delta^{++} \Delta^0 \to \Delta^{++} \Sigma^- K^+) 
&=& \sigma(\Delta^{++} \Delta^0 \to \Delta^{++} \Sigma^0 K^0) \nn \\ 
= \sigma(\Delta^{++} \Delta^- \to \Delta^+ \Sigma^- K^+) 
&=& \sigma(\Delta^{++} \Delta^- \to \Delta^+ \Sigma^0 K^0) \nn \\
= \sigma(\Delta^- \Delta^{++} \to \Delta^0 \Sigma^0 K^+)
&=& \sigma(\Delta^- \Delta^{++} \to \Delta^0 \Sigma^+ K^0) \nn \\
= \sigma(\Delta^- \Delta^+ \to \Delta^- \Sigma^0 K^+) 
&=& \sigma(\Delta^- \Delta^+ \to \Delta^- \Sigma^+ K^0), 
\\ \nn \\
\sigma(\Delta^- \Delta^0 \to \Delta^- \Sigma^- K^+)
&=& \sigma(\Delta^{++} \Delta^+ \to \Delta^{++} \Sigma^+ K^0) \nn \\
= 2\sigma(\Delta^+ \Delta^{++} \to \Delta^{++} \Sigma^0 K^+)
&=& 2\sigma(\Delta^0 \Delta^- \to \Delta^- \Sigma^0 K^0) \nn \\
= 3\sigma(\Delta^+ \Delta^+ \to \Delta^{++} \Sigma^- K^+)
&=& 3\sigma(\Delta^0 \Delta^0 \to \Delta^- \Sigma^+ K^0) \nn \\
= 6\sigma(\Delta^0 \Delta^0 \to \Delta^- \Sigma^0 K^+)
&=& 6\sigma(\Delta^+ \Delta^+ \to \Delta^{++} \Sigma^0 K^0) \nn \\
= 9\sigma(\Delta^0 \Delta^0 \to \Delta^0 \Sigma^- K^+)
&=& 9\sigma(\Delta^+ \Delta^+ \to \Delta^+ \Sigma^+ K^0) \nn \\
= 18\sigma(\Delta^+ \Delta^+ \to \Delta^+ \Sigma^0 K^+)
&=& 18\sigma(\Delta^0 \Delta^0 \to \Delta^0 \Sigma^0 K^0), 
\\ \nn \\
2\sigma(\Delta^0 \Delta^{++} \to \Delta^+ \Sigma^0 K^+)
&=& 2\sigma(\Delta^+ \Delta^- \to \Delta^0 \Sigma^0 K^0) \nn \\
= 3\sigma(\Delta^0 \Delta^+ \to \Delta^+ \Sigma^- K^+)
&=& 3\sigma(\Delta^+ \Delta^0 \to \Delta^0 \Sigma^+ K^0), 
\\ \nn \\
\sigma(\Delta^- \Delta^+ \to \Delta^0 \Sigma^- K^+)
&=& \sigma(\Delta^{++} \Delta^0 \to \Delta^+ \Sigma^+ K^0) \nn \\
= 6\sigma(\Delta^0 \Delta^+ \to \Delta^0 \Sigma^0 K^+) 
&=& 6\sigma(\Delta^+ \Delta^0 \to \Delta^+ \Sigma^0 K^0). 
\end{eqnarray}

%%%%%%%%%%%%%%%%%%%%%%%%%%%%%%%%%%%%%%%%%%%%%%%%%%%%%%%%%%%%%%%%%%%%%%%%%%
\section{Parametrizations}

The energy dependence of the total cross sections calculated so far 
will be parametrized in the following form:
\begin{equation}
\sigma (B_1 B_2 \to B_3 Y K) = a \
{\left( \frac{s}{s_0}-1 \right)}^b \
{\left( \frac{s_0}{s} \right)}^c, 
\label{paramf}
\end{equation}
where $s$ and $s_0$ are respectively the squares of the invariant collision 
energy and the threshold energy.  
$a, b$ and $c$ are the parameters to be determined 
so as to reproduce the calculated energy dependence of the total cross 
sections, with $a$ in units of $mb$.

\begin{figure}[hbt]
\epsfig{figure=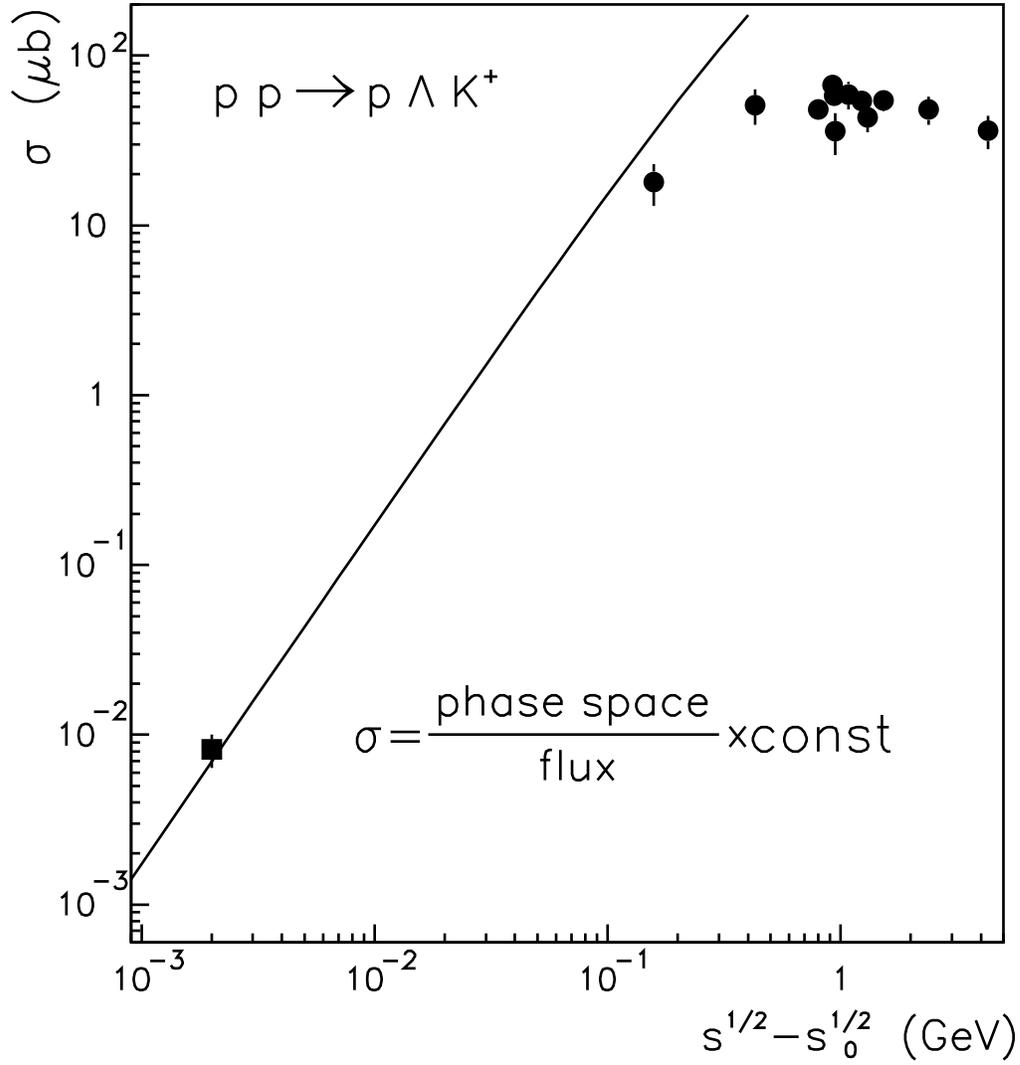,width=16cm}
\caption{\label{kt20p} Phase space consideration of the  
total cross section for the $p p \to p \Lambda K^+$ reaction,  
according to Eq.~(\protect\ref{hou}).
The dots and square show the experimental data from 
Ref.~\protect\cite{LB} and Ref.~\protect\cite{cosy}, respectively.}
\end{figure}

As was demonstrated in Ref.~\cite{many}, the functional form,  
Eq.~(\ref{paramf}), can reproduce quite well the near-threshold
behaviour of the total cross sections.
The reason is that the function reflects the energy dependence 
of the phase space reasonably just above the threshold, where the  
energy dependence of the amplitude is usually weak. In addition the 
particles in the final state may be treated nonrelativistically 
due to the small amount of energy available.
Assuming the relevant amplitude squared, $|T|^2$, 
to be a constant, the total cross section can be written as 
\begin{equation}
\label{hou}
\sigma  = \frac{R_3}{F} \  |T|^2,   
\end{equation}
with $R_3$ the three-body phase space volume and 
$F$ the flux factor. In Fig.~\ref{kt20p} we show the results for 
the energy dependence of the total cross section for the 
$p p \to p \Lambda K^+$ reaction, calculated using Eq.~(\ref{hou}), 
together with the experimental data~\cite{cosy,LB}.
The square of the constant amplitude,  
$|T|^2$, was determined so as to fit the data point of COSY~\cite{cosy}. 
The parameter, $b$, which is mostly relevant to reproduce the shape of the 
energy dependence of the total cross section near the threshold, 
is obtained to be $b=1.995$ in this case.
This value of $b$ gives a behaviour of the energy dependence  
very close to that of phase space in the nonrelativistic case~\cite{Byckling}.
On the other hand, the parameter, $b$, obtained to reproduce our results 
is, $b \simeq 2.000$, which is very close to the value of the above 
phase space calculation. The last term in Eq. (\ref{paramf}), $(s_0/s)^c$,
reflects both the energy dependence of the flux factor which becomes
important at energies $\sqrt{s}-\sqrt{s_0} > 0.5$~GeV, and also  
the contribution from meson exchange which is 
expected to behave as $s^{-2}$.

The parameters, $a$, $b$ and $c$ in Eq.~(\ref{paramf}), determined  
so as to reproduce the results calculated in our model, 
are listed in Table~\ref{param}. They are sufficient to reproduce 
all the isospin channels treated in the previous section. 
The corresponding threshold energies
squared, $s_0$, are also indicated.

%%%%%%%%%%%%%%%%%%%%%%%%%%%%%%%%%%%%%%%%%%%%%%%%%%%%%%%%%%%%%%%%%%%%%%%%%%%
\begin{table}
\caption{\label{param}
Parameterizations for energy dependence of the total cross sections.
$a$, $b$ and $c$, appearing in Eq.~(\protect\ref{paramf}) were
determined so as to reproduce the calculated total cross sections at  
energies, $\protect\sqrt{s}-\protect\sqrt{s_0} < 2$~GeV, 
with $\protect\sqrt{s}$ and $\protect\sqrt{s_0}$ being, respectively, 
the invariant collision energy and threshold energy. The parametrizations 
are recommended to be used up to invariant collision energy about 
3.6 GeV.} 
\vspace{0.5cm}
\begin{center}
\begin{tabular}{|r|l|c|c|c|c|}
\hline
No. &Reaction & $s_0$~(GeV$^2$) & $a$~(mb) & b & c \\
\hline
1 & $pp \to p \Lambda K^+ $ & 6.504 & 1.879 & 2.176 & 5.264 \\
2 & $pn \to n \Lambda K^+ $ & 6.504 & 2.812 & 2.121 & 4.893 \\
3 & $pp \to p \Sigma^0 K^+ $ & 6.904 & 5.321 & 2.753 & 8.510 \\
4 & $nn \to n \Sigma^- K^+ $ & 6.904 & 7.079 & 2.760 & 8.164 \\
5 & $pn \to n \Sigma^0 K^+ $ & 6.904 & 6.310 & 2.773 & 7.820 \\
6 & $np \to p \Sigma^- K^+$ & 6.904 & 11.02 & 2.782 & 7.674 \\
7 & $pp \to n \Sigma^+ K^+ $ & 6.904 & 1.466 & 2.743 & 3.271 \\
8 & $nn \to \Delta^- \Lambda K^+ $ & 8.085 & 6.166 & 2.842 & 1.960 \\
9 & $pp \to \Delta^{++} \Sigma^-K^+$& 8.531 & 10.00 & 2.874 & 2.543 \\
10 & $\Delta^{++} n \to p \Lambda K^+ $& 6.504 & 8.337 & 2.227 &2.511 \\
11 & $\Delta^- p \to n \Sigma^- K^+ $&6.904 &52.72 &2.799 &6.303 \\ 
12&$\Delta^{++}p \to \Delta^{++} \Lambda K^+ $ & 8.085 & 2.704 &
2.303 & 5.551 \\
13 & $\Delta^+ n \to \Delta^0 \Lambda K^+ $ & 8.085 & 0.312 &
2.110 & 2.165 \\
14 & $\Delta^+ p \to \Delta^+ \Lambda K^+ $& 8.085 & 2.917 &
2.350 & 6.557 \\
15 & $\Delta^{++} n \to \Delta^{++} \Sigma^- K^+ $ & 8.531 &
10.33 & 2.743 & 8.915 \\
16 & $\Delta^0 p \to \Delta^+ \Sigma^- K^+ $ & 8.531 & 2.128 & 
2.843 & 5.986 \\
17 & $\Delta^+ n \to \Delta^+ \Sigma^- K^+ $ & 8.531 & 10.57 &
2.757 & 10.11 \\
18 & $\Delta^{++} p \to \Delta^{++} \Sigma^0 K^+ $ & 8.531 &
10.30 & 2.748 & 9.321 \\ 
19 & $\Delta^+ n \to \Delta^0 \Sigma^0 K^+ $ & 8.531 &
1.112 & 2.846 & 5.943 \\
20 & $\Delta^+ p \to \Delta^+ \Sigma^0 K^+ $ & 8.531 &
10.62 & 2.759 & 10.20 \\
21 & $\Delta^+ p \to \Delta^0 \Sigma^+ K^+ $ & 8.531 &
0.647 & 2.830 & 3.862 \\
22 & $\Delta^+ \Delta^{++} \to \Delta^{++} \Lambda K^+ $ & 8.085 &
1.054 & 2.149 & 7.969 \\
23 & $\Delta^0 \Delta^{++} \to \Delta^+ \Lambda K^+ $ & 8.085 &
0.881 & 2.150 & 7.977 \\
24 & $\Delta^0 \Delta^+ \to \Delta^0 \Lambda K^+ $ & 8.085 &
0.291 & 2.148 & 7.934 \\
25 & $\Delta^{++} \Delta^0 \to \Delta^{++} \Sigma^- K^+ $ &8.531 &
3.532 & 2.953 & 12.06 \\
26 &$\Delta^- \Delta^0 \to \Delta^- \Sigma^- K^+ $ &8.531 &
7.047 & 2.952 & 12.05 \\
27 &$\Delta^0 \Delta^{++} \to \Delta^+ \Sigma^0 K^+ $ &8.531 & 
2.931 & 2.952 & 12.03 \\
28 &$\Delta^- \Delta^+ \to \Delta^0 \Sigma^- K^+ $ & 8.531 &
5.861 & 2.952 & 12.04 \\
\hline 
\end{tabular}
\end{center}
\end{table}
%%%%%%%%%%%%%%%%%%%%%%%%%%%%%%%%%%%%%%%%%%%%%%%%%%%%%%%%%%%%%%%%%%%%%%%%%%%%%

Note that the parameters given in 
Table~\ref{param} are determined so as to reproduce the theoretical 
results at energies up to about 2~GeV above the 
corresponding threshold energies.
The reason is, when kaon production is studied at energies 
larger than 2~GeV above threshold, we need to include more resonances 
which decay to kaon and hyperon in the model, as well as taking 
into account the other kaon production processes with more than three 
particles appearing in the final states, 
where it seems impossible due to the present experimental data
available. In practice, we believe that the parametrizations are 
recommended to be used up to invariant collision energies about 
3.6~GeV, which are usually enough.

Although we have parametrized explicitly all the   
isospin channels required, it has been traditional to use isospin-averaged 
cross sections. In order to compare our results with those 
isospin-averaged cross sections, we have also calculated  
them within the model.
The results are again parametrized by the function, Eq.~(\ref{paramf}). 
The parameters, $a, b$ and $c$, for the isospin-averaged cross sections, 
are listed in Table~\ref{param1}.
Note that the parameterizations in Table~\ref{param1} 
also include $K^0$ production channels.
Thus, the parametrizations for the $K^+$ production channels alone 
can be obtained replacing the values for the parameter $a$ in
Table~\ref{param1} by $\frac{1}{2} a$.

%
%%%%%%%%%%%%%%%%%%%%%%%%%%%%%%%%%%%%%%%%%%%%%%%%%%%%%%%%%%%%%%%%%%%%%%%%%%%%
\begin{table}
\caption{\label{param1}
Parameterizations for the energy dependence of the isospin-averaged 
total cross sections. (See also the caption of Table~\protect\ref{param}.) 
For the $K^+$ production reactions, replace the values for the parameter 
$a$ by $\frac{1}{2} a$.}
\vspace{0.5cm}
\begin{center}
\begin{tabular}{|r|l|c|c|c|}
\hline
No. &Reaction & $a$~(mb) & b & c \\
\hline
1 & $N N \to N \Lambda K$ & 2.330 & 2.140 & 5.024 \\
2 & $N N \to N \Sigma K$ & 15.49 & 2.768 & 7.222 \\
3 & $N N \to \Delta \Lambda K$ & 9.249 & 2.842 & 1.960 \\
4 & $N N \to \Delta \Sigma K$ & 12.50 & 2.874 & 2.543 \\
5 & $ \Delta N \to N \Lambda K $ & 4.169 & 2.227 & 2.511 \\
6 & $\Delta N \to N \Sigma K $ & 39.54 & 2.799 & 6.303 \\
7 & $\Delta N \to \Delta \Lambda K$ & 2.679 & 2.280 & 5.086 \\
8 & $\Delta N \to \Delta \Sigma K$ &21.18 & 2.743 & 8.407 \\
9 & $\Delta \Delta \to \Delta \Lambda K$ & 0.337 & 2.149 & 7.967 \\
10&$\Delta \Delta \to \Delta \Sigma K$ &5.140 & 2.952 & 12.05 \\
\hline
\end{tabular}
\end{center}
\end{table}
%%%%%%%%%%%%%%%%%%%%%%%%%%%%%%%%%%%%%%%%%%%%%%%%%%%%%%%%%%%%%%%%%%%%%%%%%
%

\begin{figure}[hbt]
\epsfig{figure=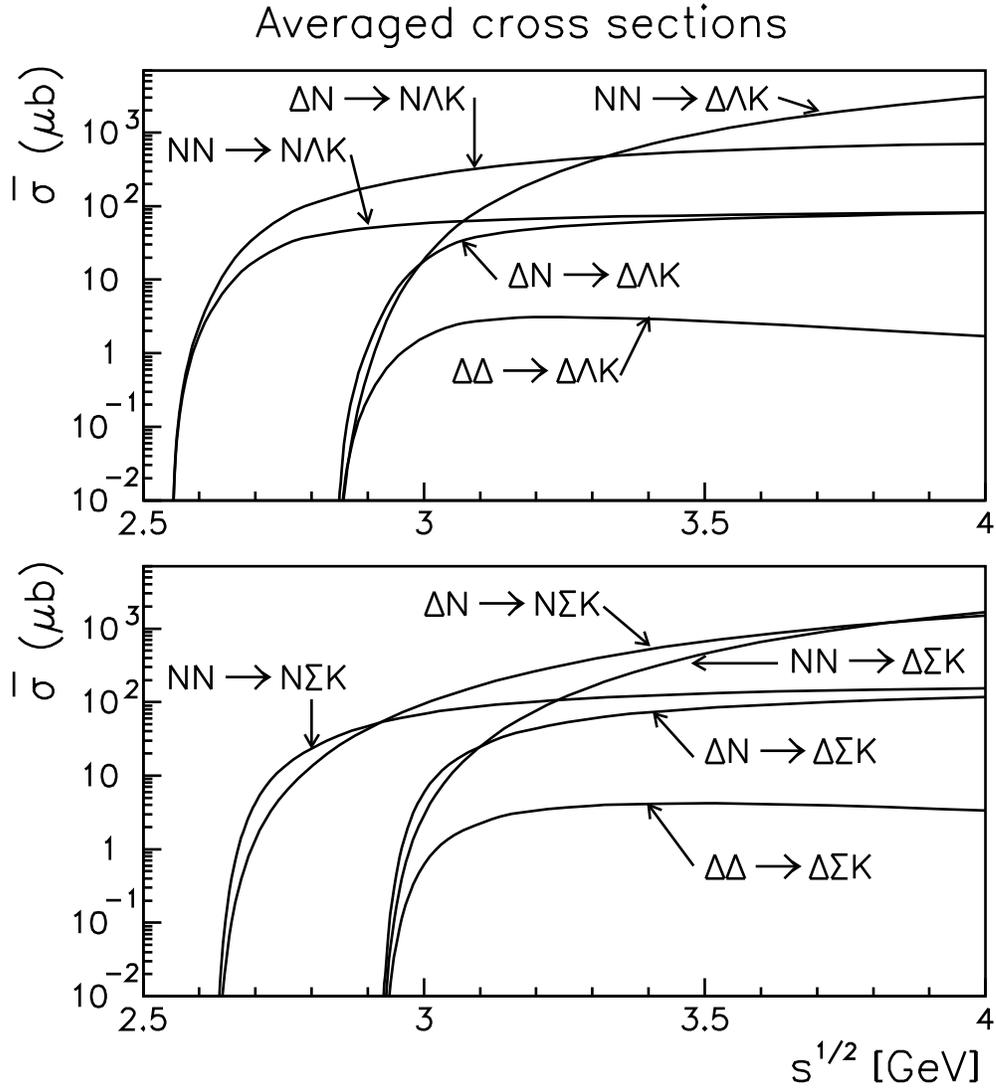,width=16cm}
\caption{\label{kt26p} Energy dependence of the isospin-averaged 
total cross sections for the $\Lambda$ and 
$\Sigma$ production reactions in baryon baryon reactions.}
\end{figure}

We show the energy dependence of the isospin-averaged total  
cross sections, $\bar{\sigma}$, calculated in the model for all
the baryon baryon reactions in Fig.~\ref{kt26p}.
Note that the horizontal axes indicate invariant collision energies.
Our results illustrate that the magnitudes of the isospin-averaged 
cross sections at energies, $\sqrt{s}-\sqrt{s_0} < 200$~MeV, 
are, $\bar{\sigma}(\Delta N \to N Y K) \simeq 
\bar{\sigma}(N N \to N Y K)$ for both the $\Lambda$ and 
$\Sigma$ production channels.  
However, at energies, $\sqrt{s}-\sqrt{s_0} > 200$~MeV, the magnitudes 
of the cross sections $\bar{\sigma}(\Delta N \to N Y K)$ 
become about 6 times larger  
than those of $\bar{\sigma}(N N \to \Delta Y K)$.
It is also noticable that the magnitudes of 
$\bar{\sigma}(N N \to \Delta Y K)$ become comparable with those 
of $\bar{\sigma}(N N \to N Y K)$ at $N N$ invariant energies about 3.1 GeV, 
and overcome factors of 5 to 10 at $N N$ invariant energies about 3.5 GeV.
In particular, this is pronounced for the 
$\Lambda$ production reactions. It is not clear at the moment   
how much the $N N \to \Delta Y K$ reactions contribute    
to the total kaon yield in heavy ion collisions because they 
have usually been neglected. It seems worthwhile to perform 
calculations including these reactions. 
At invariant energies larger than about 3.2~GeV, the magnitudes
of $\bar{\sigma}(\Delta N \to \Delta Y K)$ are also comparable 
to those of $\bar{\sigma}(N N \to N Y K)$.

It is interesting to see if the relation for the 
elementary kaon production cross sections,  
$\bar{\sigma}(N N \to \Delta \Lambda K) >
\bar{\sigma}(\Delta N \to N \Lambda K) >
\bar{\sigma}(N N \to N \Lambda K)$, can   
appreciable influence inclusive kaon production 
in proton-nucleus or heavy ion collisions.
If this is the case, high energy 
inclusive kaon production in heavy ion collisions will be also  
significantly influenced by the $\Delta$.
A test for this argument can be made by performing  
calculations for high energy inclusive kaon production  
both, neglecting all the $\Delta$'s, and including them, 
and comparing the results with the existing experimental data.
 
Next, we compare the isospin-averaged total cross sections 
calculated in this model, with the commonly used relation
suggested by Randrup and Ko~\cite{ran}. 
For the $N N \to N Y K$ reactions, they suggested the relations:
\bg
{\bar \sigma}(NN \to N \Lambda K) &=& 3 \ \sigma(pp\to p\Lambda K^+), 
\label{pre1}\\
{\bar \sigma}(NN \to N \Sigma K) &=& 3 \ \left[\sigma(pp\to p\Sigma^0 K^+)+
\sigma(pp\to p\Sigma^+ K^0)\right].
\label{pre2}
\en

\begin{figure}[hbt]
\epsfig{figure=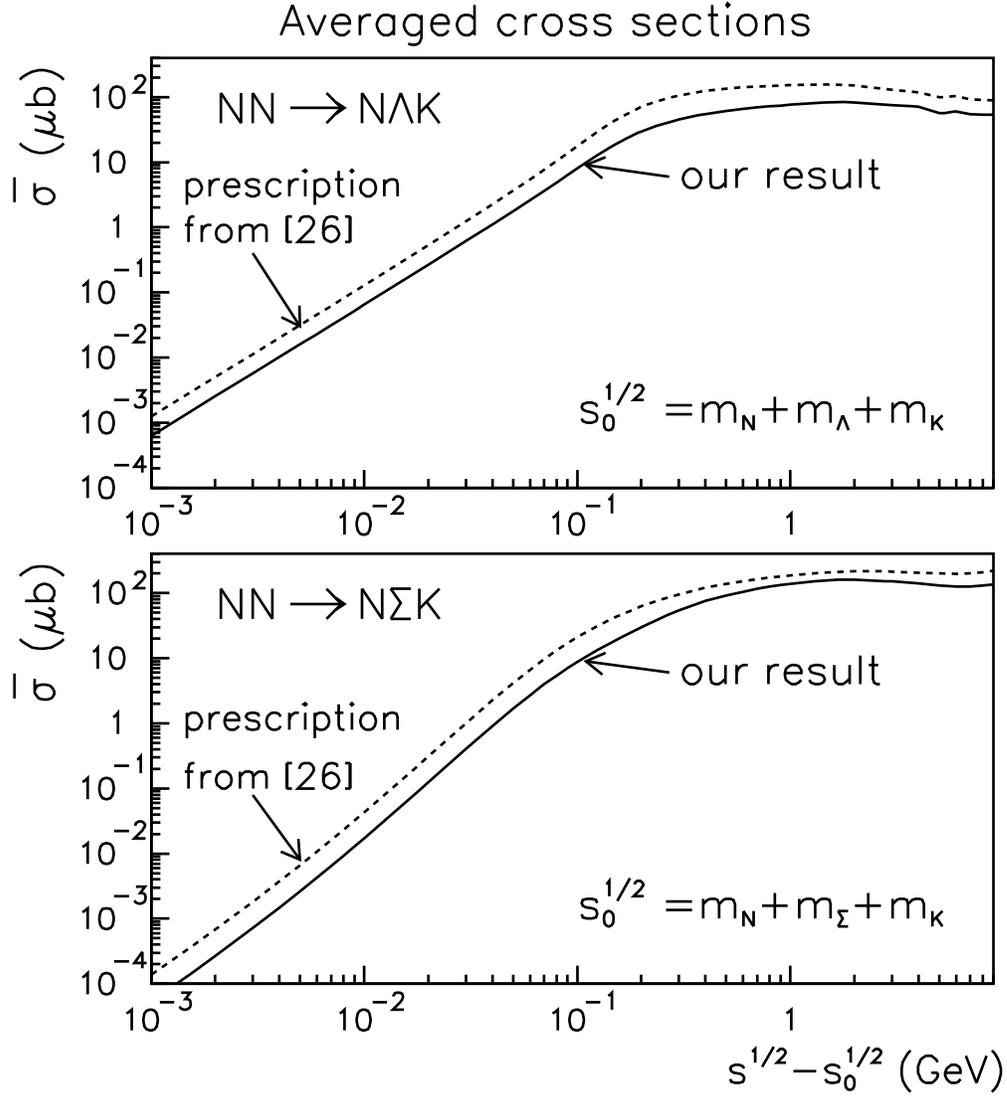,width=16cm}
\caption{\label{kt24p} Energy dependence of the total cross sections 
both calculated in the model, isospin-averaged, and 
using the relations Eqs.~(\protect\ref{pre1}) 
and~(\protect\ref{pre2}) suggested by Randrup and Ko~\protect\cite{ran}.
The solid lines show the model calculations for the isospin-averaged 
total cross sections (denoted by "our result"), 
while the dashed lines show the model calculations   
obtained using the right hand side of Eqs.~(\protect\ref{pre1}) 
and~~(\protect\ref{pre2}) 
(denoted by "prescription from~\protect\cite{ran}").} 
\end{figure}

In Fig.~\ref{kt24p} we show the isospin-averaged total cross sections 
calculated in the model, $\bar{\sigma}(NN \to N \Lambda K)$ and 
$\bar{\sigma}(NN \to N \Sigma K)$ (denoted by our model), 
and the quantities calculated  
using the right hand side of Eqs.~(\ref{pre1}) and~(\ref{pre2}) 
(denoted by prescription from~\cite{ran}).  
The explicit calculation shows disagreement of about a factor 2 
between the relation suggested by 
Randrup and Ko~\cite{ran} for most of the energy range up to 2 GeV 
above the thresholds, for both the $\Lambda$ and $\Sigma$ 
production reactions. Thus, the calculations performed using the relation
suggested by Randrup and Ko~\cite{ran} to obtain the isospin-averaged 
cross sections for the $N N \to N Y K$ reactions, may have about a       
factor of 2 ambiguity for these reactions.

As for the $\Delta N \to N Y K$ reactions, they 
suggested the relations~\cite{ran}:
\bg
{\bar \sigma}(\Delta N \to N \Lambda K) &=&
\frac{9}{4} \ \sigma(pp\to p\Lambda K^+),
\label{pre3}\\
{\bar \sigma}( \Delta N \to N \Sigma K) &=&
\frac{9}{4} \ \left[\sigma(pp\to p\Sigma^0 K^+)+
\sigma(pp\to p\Sigma^+ K^0)\right].
\label{pre4}
\en

\begin{figure}[hbt]
\epsfig{figure=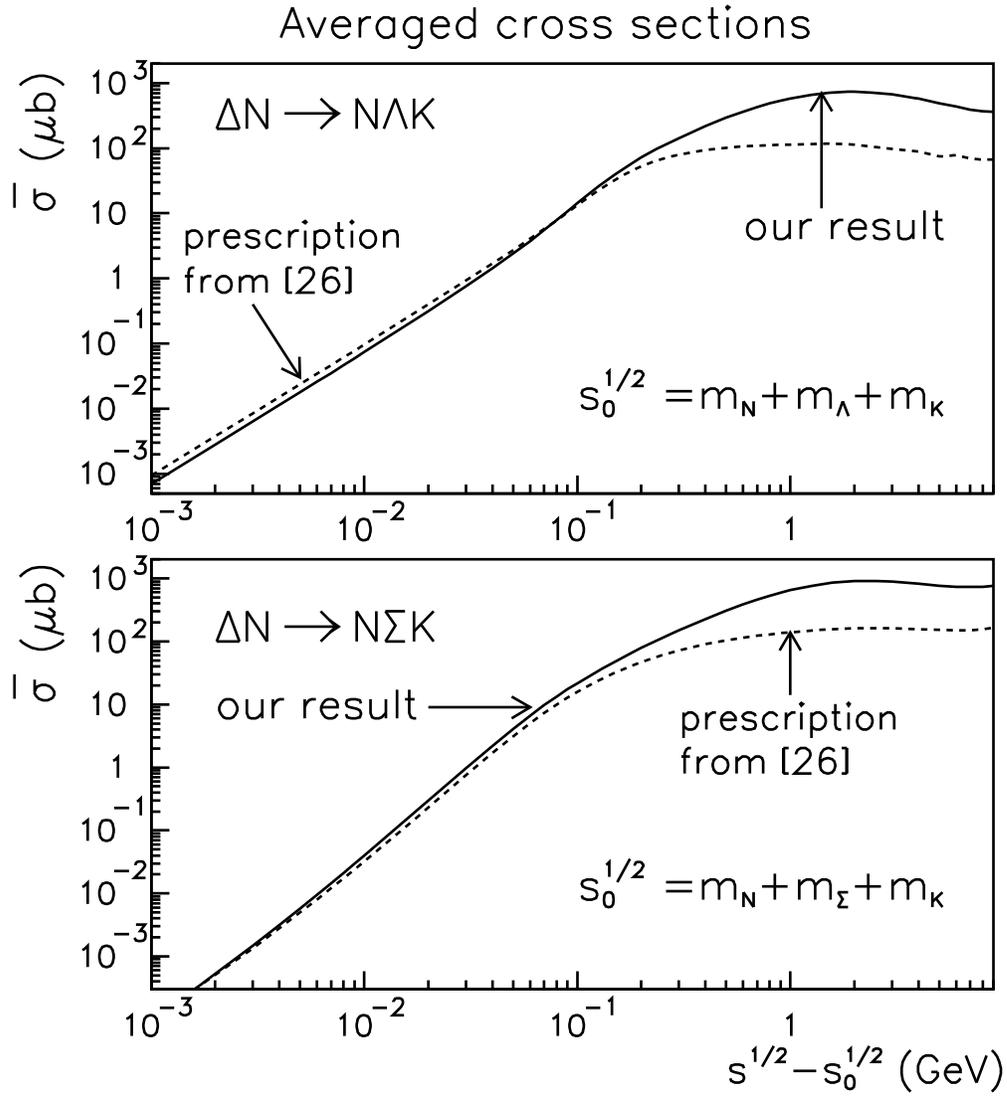,width=16cm}
\caption{\label{kt25p} Same as in Fig.~\protect\ref{kt24p}, but
for the $\Delta N \to N \Lambda K$ and $\Delta N \to N \Sigma K$ 
reactions and Eqs.~(\protect\ref{pre3}) and~~(\protect\ref{pre4}),
respectively.} 
\end{figure}

In Fig.~\ref{kt25p} we show similar quantities calculated 
within the present model as for the $N N \to N Y K$ reactions.
At energies near the thresholds, $\sqrt{s}-\sqrt{s_0} < 200$~MeV,
our results are in good agreement 
with the relations Eqs.~(\ref{pre3}) and (\ref{pre4}), suggested 
to obtain the isospin-averaged cross sections.
However, at energies, $\sqrt{s} - \sqrt{s_0} \simeq 1$~GeV, 
our results show a factor of 6 discrepancy with 
the relations suggested by Randrup and Ko~\cite{ran}.

It is worth noting that the recent calculation for kaon production 
in heavy ion collisions in free space~\cite{Cassingn},  
performed at energies near or below  
the threshold region of the $N N \to N \Lambda K$ reaction,  
illustrates that the dominant contribution 
comes from the secondary, pion induced reactions.   
Cassing et al.~\cite{Cassingn} and Li and 
Ko~\cite{libbk} used the relations of 
Eqs.~(\ref{pre1}) --~(\ref{pre4}), 
to obtain the isospin-averaged cross sections. 
In view of our results, it seems necessary to perform calculations 
using the total cross sections parametrized 
in a consistent manner for both the baryon baryon and pion baryon 
reactions. 

\section{Application to heavy-ion collisions}

Kaon production in heavy-ion collisions at SIS energies (1-2 AGeV)
has been studied theoretically using various transport models.
Important inputs for such studies are the elementary 
kaon production cross sections in pion baryon and baryon baryon
interactions. Conclusions concerning nuclear matter equations
of state and kaon in-medium properties, obtained by comparing
calculations with measurements, depend on the input kaon production
cross sections. 

In Ref. \cite{lilee}, a detailed study of these elementary cross
sections was carried out, based on both the parametrization of
experimental data and theoretical model calculations. Here we want
to compare the kaon yields in heavy-ion collisions obtained
using the cross sections in this work with the cross sections
of Ref. \cite{lilee}. We will use the same relativistic transport
model as in Ref. \cite{lilee} so that the difference
in kaon yield is due entirely to the different elementary cross
sections used. In this way, we can get a feeling for the
sensitivity of transport model predictions to the
underlying elementary cross sections.

\begin{figure}[hbt]
\epsfig{figure=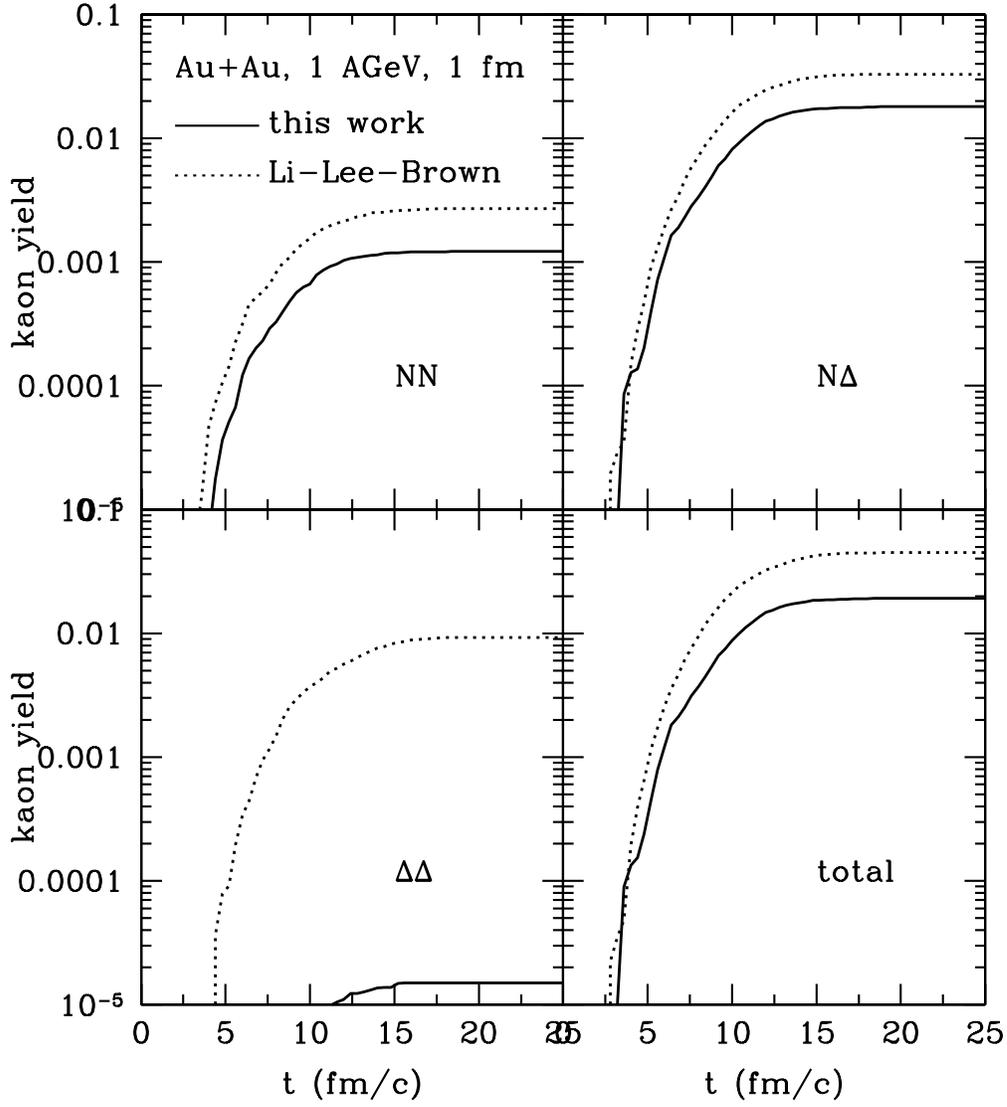,width=16cm}
\caption{\label{auau} 
Kaon yield in Au+Au collisions at 1 AGeV and 1 fm obtained
with two sets of elementary cross sections.}
\end{figure}

\begin{figure}[hbt]
\epsfig{figure=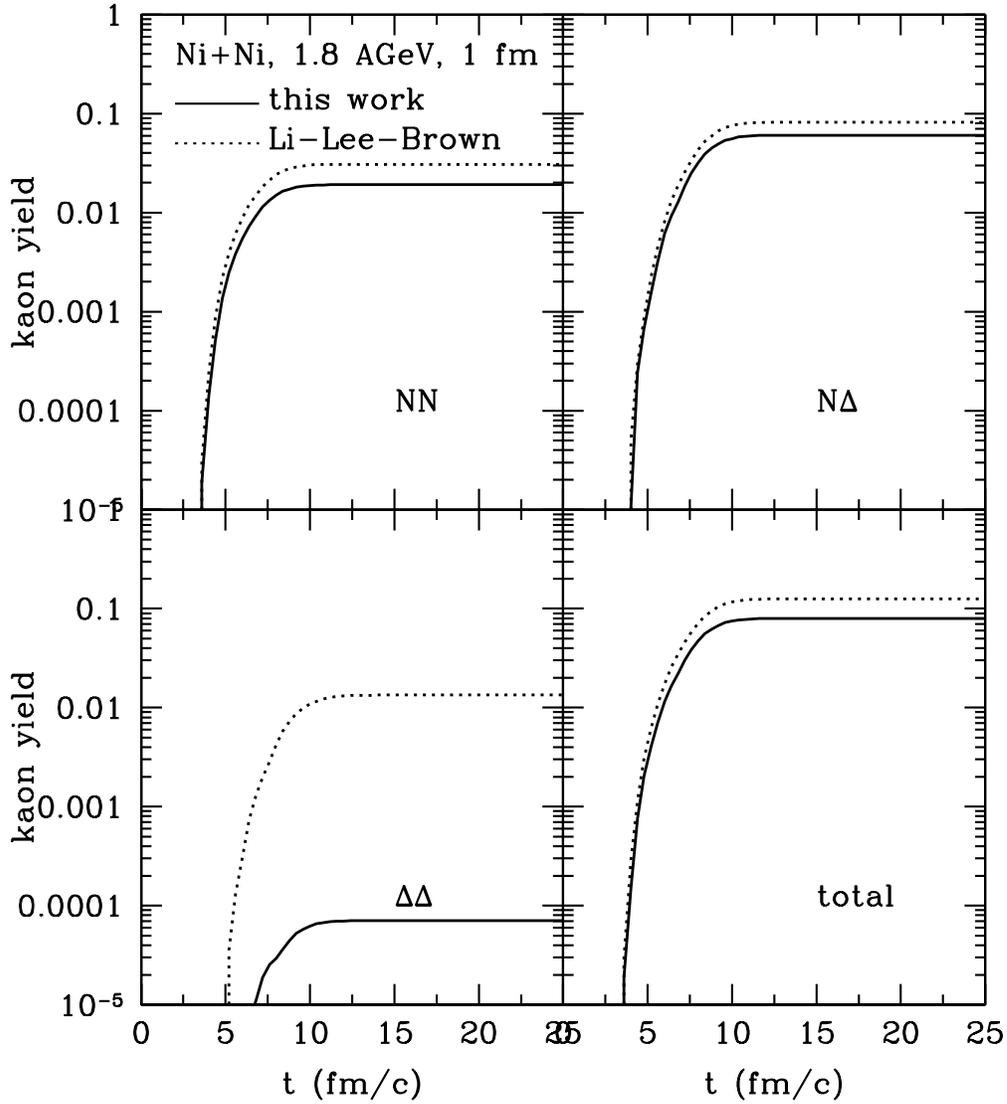,width=16cm}
\caption{\label{nini} 
The same as Fig. \protect\ref{auau}, for Ni+Ni collisions
at 1.8 AGeV.}
\end{figure}

We take central Au+Au collsions at 1 AGeV and Ni+Ni collisions
at 1.8 AGeV as examples. The results are shown in 
Figs. \ref{auau} and \ref{nini}, respectively. From these 
figures, we can make the following observations:

\begin{enumerate}

\item Generally, the kaon yield obtained with the cross sections
from this work is smaller than that obtained with cross
sections from Ref. \cite{lilee}. The difference is about a
factor of two at 1 AGeV, and reduces to about 40\% at
1.8 AGeV.

\item Using both sets of cross sections, the $N\Delta$ channel
is found to make the largest contributions among all the
baryon baryon interactions. This is also in agreement with
the earlier findings using the simple Randrup-Ko 
parametrization. 

\item The contribution from the $\Delta\Delta$ channel
is particularly small using the cross sections from this
work. This is due to the fact that in the present model, the
cross section for $\Delta\Delta \rightarrow NYK$ channel
is zero, while in Ref. \cite{lilee} off-shell pion exchange
was included and gave a small contribution.

\end{enumerate}

%
%%%%%%%%%%%%%%%%%%%%%%%%%%%%%%%%%%%%%%%%%%%%%%%%%%%%%%%%%%%%%%%%%%%%%%%%%
\section{Summary}

Within the resonance model, which is fully relativistic, but includes 
only the tree-level processes as an effective and empirical contributions,  
we have studied the elementary kaon 
production reactions in baryon baryon collisions in a consistent 
manner with those for the pion baryon reactions. We have calculated 
and parametrized the energy dependence of the total cross sections for 
all the necessary isospin channels in the 
$N N \to N Y K$, $N N \to \Delta Y K$, $\Delta N \to N Y K$,
$\Delta N \to \Delta Y K$ and $\Delta \Delta \to \Delta Y K$
reactions ($Y=\Lambda$ and $\Sigma$). 
We have also calculated and parametrized those for the isospin-averaged
cross sections and compared with the the relation suggested by Rundrup and 
Ko~\cite{ran} to obtain the isospin-averaged cross sections.

The present study has several points which may need to be improved,  
for example, consistency with the unitarity, the possibility to add other  
types of meson exchanges, such as kaon exchange, and the overestimate for 
the $p n (n p) \to N Y K$ reactions and the reactions 
involving the $\Delta$, we would like to emphasize 
a practical aspect of this work, which we belive providing a useful inputs 
for further investigations of kaon production in heavy ion collisions. 
Keeping these in mind, we state a few comments as follows.
First, our explicit calculation indicates that   
the contributions from the $\Delta N \to N Y K$ and 
$N N \to \Delta Y N$ reactions are substantially larger than those
from the $N N \to N Y K$ reactions at higher energies. 
Second, the relation suggested by Randrup and Ko~\cite{ran}
(c.f. Eqs.~(\ref{pre1}) and~(\ref{pre2})) 
deviates from our results by about a factor of 2 for the 
$N N \to N Y K$ reactions. 
For the $\Delta N \to N Y K$ reactions, the relation 
(c.f. Eqs.~(\ref{pre3}) and~(\ref{pre4})) holds 
well at low energies, up to about 200 MeV above the threshold, 
but deviations by as much as a factor of 6 have emerged at higher 
energies. From these facts we conclude that it is necessary 
to use consistent parametrizations of  
total kaon production cross sections for both the pion baryon 
and baryon baryon reactions. Furthermore, one should use 
isospin-averaged total cross sections 
calculated in a consistent manner.  
Lastly although the present study has been made for exclusive kaon 
production, there might also be a significant difference between 
the $\Delta N$ and $N N$ collision reactions in  
inclusive kaon production in proton-nucleus or heavy ion collisions.
Thus, it is interesting to perform calculations for inclusive kaon   
production, focusing on the effect of the $\Delta$. \\ 

\noindent
{\bf Acknowledgement}\\
We thank J. Haidenbauer for helpful comments on the unitarity 
and coupled channel calculation.
K.T. would like to thank K. Yazaki for valuable discussions at 
the very early stage of this work. 
A.S. would like to thank W. Cassing and U. Mosel for valuable
discussions. This work was supported in part by the Forschungszentrum
J\"{u}lich, the Australian Research Council, and the U.S. Department
of Energy under grant No. DE-FG02-88ER40388.
\\ \\ \\

%%%%%%%%%%%%%%%%%%%%%%%%%%%%%%%%%%%%%%%%%%%%%%%%%%%%%%%%%%%%%%%%%%%%%%%%
\section{Appendix}

\subsection{Coupling constants of the resonances}

Here we give relations between the branching ratios of the 
adopted resonances and the corresponding coupling constants 
squared, evaluated in the rest frame of each resonance.
Note that all the coupling constants appearing below should be  
understood as being multiplied by the appropriate form factors.
In addition, the confidence level, spin and parity of each resonance 
is specified. For the definition of $\lambda(x,y,z)$ appearing below, 
see the text, after Eq.~(\ref{crosssection}).

%
%%%%%%%%%%%%%%%%%%%%%%%%%%%%%%%%%%%%%%%%%%%%%%%%%%%%%%%%%%%%%%%%%%%%%%%%
\subsubsection{$N(1650)^{****}(\frac{1}{2})^-$}
\bg
\Gamma(N(1650) \rightarrow N \pi) 
&=& 3 \frac{g^2_{\pi N N(1650)} }{4\pi} 
\frac{(E_N + m_N)}{m_{N(1650)}} |{\vec p}_N|, \nn
\\
{\rm with}\quad  |{\vec p}_N| &=& \frac{\lambda^{1/2}
(m_{N(1650)}^2,m_N^2,m_\pi^2)}{2m_{N(1650)}},
\label{anp}
\\
\Gamma(N(1650) \rightarrow N \eta)
&=& \frac{g^2_{\eta N N(1650)} }{4\pi} 
\frac{(E_N + m_N)}{m_{N(1650)}} |{\vec p}_N|, \nn
\\
{\rm with}\quad  |{\vec p}_N| &=& \frac{\lambda^{1/2}
(m_{N(1650)}^2,m_N^2,m_\eta^2)}{2m_{N(1650)}},
\label{ane}
\\
\Gamma(N(1650) \rightarrow \Delta \pi) 
&=& 2 \frac{g^2_{\pi \Delta N(1650)}}{6\pi}
\frac{m_{N(1650)} (E_\Delta - m_\Delta)}{m_\pi^2 m_\Delta^2}
|{\vec p}_\Delta|^3, \nn 
\\
{\rm with}\quad |{\vec p}_\Delta| &=& \frac{\lambda^{1/2}
(m_{N(1650)}^2,m_\Delta^2,m_\pi^2)}{2m_{N(1650)}}, 
\label{aip}
\\
\Gamma(N(1650) \rightarrow \Lambda K) 
&=& \frac{g^2_{\Lambda K N(1650)}}{4\pi}
\frac{(E_\Lambda + m_\Lambda)}{m_{N(1650)}} |{\vec p}_\Lambda|, \nn
\\
{\rm with}\quad |{\vec p}_\Lambda| &=& \frac{\lambda^{1/2}
(m_{N(1650)}^2,m_\Lambda^2,m_K^2)}{2m_{N(1650)}}.
\label{alk}
\en
%%%%%%%%%%%%%%%%%%%%%%%%%%%%%%%%%%%%%%%%%%%%%%%%%%%%%%%%%%
\subsubsection{$N(1710)^{***}(\frac{1}{2})^+$}
\bg
\Gamma(N(1710) \rightarrow N \pi)
&=& 3 \frac{g^2_{\pi N N(1710)} }{4\pi}
\frac{(E_N - m_N)}{m_{N(1710)}} |{\vec p}_N|, \nn
\\
{\rm with}\quad  |{\vec p}_N| &=& \frac{\lambda^{1/2}
(m_{N(1710)}^2,m_N^2,m_\pi^2)}{2m_{N(1710)}},
\label{bnp}
\\
\Gamma(N(1710) \rightarrow N \eta)
&=& \frac{g^2_{\eta N N(1710)} }{4\pi}
\frac{(E_N - m_N)}{m_{N(1710)}} |{\vec p}_N|, \nn
\\
{\rm with}\quad  |{\vec p}_N| &=& \frac{\lambda^{1/2}
(m_{N(1710)}^2,m_N^2,m_\eta^2)}{2m_{N(1710)}},
\label{bne}
\\
\Gamma(N(1710) \rightarrow N \rho)
&=& 3 \frac{g^2_{\rho N N(1710)} }{4\pi}
\frac{|{\vec p}_N|}{m_{N(1710)}} 
\left[ E_N - 3m_N + E_\rho 
(\frac{m^2_{N(1710)}-m^2_N}{m^2_\rho} - 1) \right], \nn
\\
{\rm with}\quad  |{\vec p}_N| &=& \frac{\lambda^{1/2}
(m_{N(1710)}^2,m_N^2,m_\rho^2)}{2m_{N(1710)}},
\label{bnr}
\\
\Gamma(N(1710) \rightarrow \Delta \pi)
&=& 2 \frac{g^2_{\pi \Delta N(1710)}}{6\pi}
\frac{m_{N(1710)} (E_\Delta + m_\Delta)}{m_\pi^2 m_\Delta^2}
|{\vec p}_\Delta|^3, \nn
\\
{\rm with}\quad |{\vec p}_\Delta| &=& \frac{\lambda^{1/2}
(m_{N(1710)}^2,m_\Delta^2,m_\pi^2)}{2m_{N(1710)}},
\label{bip}
\\
\Gamma(N(1710) \rightarrow \Lambda K)
&=& \frac{g^2_{\Lambda K N(1710)}}{4\pi}
\frac{(E_\Lambda - m_\Lambda)}{m_{N(1710)}} |{\vec p}_\Lambda|, \nn
\\
{\rm with}\quad |{\vec p}_\Lambda| &=& \frac{\lambda^{1/2}
(m_{N(1710)}^2,m_\Lambda^2,m_K^2)}{2m_{N(1710)}},
\label{blk}
\\
\Gamma(N(1710) \rightarrow \Sigma K)
&=& 3 \frac{g^2_{\Sigma K N(1710)}}{4\pi}
\frac{(E_\Sigma - m_\Sigma)}{m_{N(1710)}} |{\vec p}_\Sigma|, \nn
\\
{\rm with}\quad |{\vec p}_\Sigma| &=& \frac{\lambda^{1/2}
(m_{N(1710)}^2,m_\Sigma^2,m_K^2)}{2m_{N(1710)}}. 
\label{bsk}
\en
%%%%%%%%%%%%%%%%%%%%%%%%%%%%%%%%%%%%%%%%%%%%%%%%%%%%%%%%%%
\subsubsection{$N(1720)^{****}(\frac{3}{2})^+$}
\bg
\Gamma(N(1720) \rightarrow N \pi)
&=& 3 \frac{g^2_{\pi N N(1720)} }{12\pi}
\frac{(E_N + m_N)}{m_{N(1720)} m_\pi^2} |{\vec p}_N|^3, \nn
\\
{\rm with}\quad  |{\vec p}_N| &=& \frac{\lambda^{1/2}
(m_{N(1720)}^2,m_N^2,m_\pi^2)}{2m_{N(1720)}},
\label{cnp}
\\
\Gamma(N(1720) \rightarrow N \eta)
&=& \frac{g^2_{\pi N N(1720)}}{12\pi}
\frac{(E_N + m_N)}{m_{N(1720)} m_\eta^2} |{\vec p}_N|^3, \nn
\\
{\rm with}\quad  |{\vec p}_N| &=& \frac{\lambda^{1/2}
(m_{N(1720)}^2,m_N^2,m_\eta^2)}{2m_{N(1720)}},
\label{cne}
\\
\Gamma(N(1720) \rightarrow N \rho)
&=& 3 \frac{g^2_{\rho N N(1720)}}{12\pi}
\frac{(E_N - m_N)}{m_{N(1720)}} 
\left[ 3 + \frac{\vecp_N^{\, 2}}{m_\rho^2} \right]
|{\vec p}_N|, \nn
\\
{\rm with}\quad  |{\vec p}_N| &=& \frac{\lambda^{1/2}
(m_{N(1720)}^2,m_N^2,m_\rho^2)}{2m_{N(1720)}},
\label{cnr}
\\
\Gamma(N(1720) \rightarrow \Delta \pi)
&=& 2 \frac{g^2_{\pi \Delta N(1720)}}{36\pi}
\frac{m_\Delta |\vecp_\Delta|}{m_{N(1720)}} \left[\, 
(\frac{E_\Delta}{m_\Delta}) - 1\,\right]
\left[\,2(\frac{E_\Delta}{m_\Delta})^2 -
2(\frac{E_\Delta}{m_\Delta}) + 5\,\right] , \nn
\\
{\rm with}\quad  |{\vec p}_\Delta| &=& \frac{\lambda^{1/2}
(m_{N(1720)}^2,m_\Delta^2,m_\pi^2)}{2m_{N(1720)}},
\label{cip}
\\
\Gamma(N(1720) \rightarrow \Lambda K)
&=& \frac{g^2_{K \Lambda N(1720)} }{12\pi}
\frac{(E_\Lambda + m_\Lambda)}{m_{N(1720)} m_K^2} |{\vec p}_\Lambda|^3, \nn
\\
{\rm with}\quad  |{\vec p}_\Lambda| &=& \frac{\lambda^{1/2}
(m_{N(1720)}^2,m_\Lambda^2,m_K^2)}{2m_{N(1720)}},
\label{clk}
\\
\Gamma(N(1720) \rightarrow \Sigma K)
&=& 3 \frac{g^2_{K \Sigma N(1720)} }{12\pi}
\frac{(E_\Sigma + m_\Sigma)}{m_{N(1720)} m_K^2} |{\vec p}_\Sigma|^3, \nn
\\
{\rm with}\quad  |{\vec p}_\Sigma| &=& \frac{\lambda^{1/2}
(m_{N(1720)}^2,m_\Sigma^2,m_K^2)}{2m_{N(1720)}}.
\label{csk}
\en
%%%%%%%%%%%%%%%%%%%%%%%%%%%%%%%%%%%%%%%%%%%%%%%%%%%%%%%%%%
\subsubsection{$\Delta(1920)^{***}(\frac{3}{2})^+$}
\bg
\Gamma( \Delta(1920) \rightarrow N \pi)
&=& \frac{g^2_{\pi N \Delta(1920)} }{12\pi}
\frac{(E_N + m_N)}{m_{\Delta(1920)} m_\pi^2} |{\vec p}_N|^3, \nn
\\
{\rm with}\quad  |{\vec p}_N| &=& \frac{\lambda^{1/2}
(m_{\Delta(1920)}^2,m_N^2,m_\pi^2)}{2m_{\Delta(1920)}},
\label{dnp}
\\
\Gamma( \Delta(1920) \rightarrow \Sigma K)
&=& \frac{g^2_{\pi N \Delta(1920)} }{12\pi}
\frac{(E_\Sigma + m_\Sigma)}{m_{\Delta(1920)} m_K^2} 
|{\vec p}_\Sigma|^3, \nn
\\
{\rm with}\quad  |{\vec p}_\Sigma| &=& \frac{\lambda^{1/2}
(m_{\Delta(1920)}^2,m_\Sigma^2,m_K^2)}{2m_{\Delta(1920)}}.
\label{dsk}
\en
%%%%%%%%%%%%%%%%%%%%%%%%%%%%%%%%%%%%%%%%%%%%%%%%%%%%%%%%%%%%%%%%%%%%%%%%%

\subsection{Cross section relation for the $\pi N \to \Lambda K$ reactions}

Here, we add a relation for the total cross sections for the 
$\pi N \to \Lambda K$ reactions, 
which was not mentioned in Ref.~\cite{tsu}. The relation is:
\bg
    \sigma(\pi^+ n \to \Lambda K^+) &=& \sigma(\pi^- p \to \Lambda K^0) 
\nn \\
= 2 \sigma(\pi^0 p \to \Lambda K^+) &=& 2 \sigma(\pi^0 n \to \Lambda K^0).
\label{pin}
\en
%

%%%%%%%%%%%%%%%%%%%%%%%%%%%%%%%%%%%%%%%%%%%%%%%%%%%%%%%%%%%%%%%%%%%%%%%%
%References

%%%%%%%%%%%%%%%%%%%%%%%%%%%%%%%%%%%%%%%%%%%%%%%%%%%%%%%%%%%%%%%%%%%

%\end{document}
%%%% Figures
%%%%%%%%%%%%%%%%%%%%%%%%%%%%%%%%%%%%%%%%%%%%%%%%%%%%%%%%%%%%%%%%%%
\newpage
%%%%%%%%%%%%%%%%%%%%%%%%%%%%%%%%%%%%%%%%%%%%%%%%%%%%%%%%%%%%%%%%%%
\newpage
%%%%%%%%%%%%%%%%%%%%%%%%%%%%%%%%%%%%%%%%%%%%%%%%%%%%%%%%%%%%%%%%%
\newpage
%%%%%%%%%%%%%%%%%%%%%%%%%%%%%%%%%%%%%%%%%%%%%%%%%%%%%%%%%%%%%%%%%
\newpage
%%%%%%%%%%%%%%%%%%%%%%%%%%%%%%%%%%%%%%%%%%%%%%%%%%%%%%%%%%%%%%%%%
\newpage
%%%%%%%%%%%%%%%%%%%%%%%%%%%%%%%%%%%%%%%%%%%%%%%%%%%%%%%%%%%%%%%%%
\newpage
%%%%%%%%%%%%%%%%%%%%%%%%%%%%%%%%%%%%%%%%%%%%%%%%%%%%%%%%%%%%%%%%%
\newpage
\clearpage
%%%%%%%%%%%%%%%%%%%%%%%%%%%%%%%%%%%%%%%%%%%%%%%%%%%%%%%%%%%%%%%%%
\newpage
%%%%%%%%%%%%%%%%%%%%%%%%%%%%%%%%%%%%%%%%%%%%%%%%%%%%%%%%%%%%%%%%%
\newpage
%%%%%%%%%%%%%%%%%%%%%%%%%%%%%%%%%%%%%%%%%%%%%%%%%%%%%%%%%%%%%%%%%
\newpage
%%%%%%%%%%%%%%%%%%%%%%%%%%%%%%%%%%%%%%%%%%%%%%%%%%%%%%%%%%%%%%%%%
\newpage
%%%%%%%%%%%%%%%%%%%%%%%%%%%%%%%%%%%%%%%%%%%%%%%%%%%%%%%%%%%%%%%%%
\newpage
%%%%%%%%%%%%%%%%%%%%%%%%%%%%%%%%%%%%%%%%%%%%%%%%%%%%%%%%%%%%%%%%%
\newpage
%%%%%%%%%%%%%%%%%%%%%%%%%%%%%%%%%%%%%%%%%%%%%%%%%%%%%%%%%%%%%%%%%
%\newpage
%%%%%%%%%%%%%%%%%%%%%%%%%%%%%%%%%%%%%%%%%%%%%%%%%%%%%%%%%%%%%%%%%
\newpage
%%%%%%%%%%%%%%%%%%%%%%%%%%%%%%%%%%%%%%%%%%%%%%%%%%%%%%%%%%%%%%%%%
\newpage
%%%%%%%%%%%%%%%%%%%%%%%%%%%%%%%%%%%%%%%%%%%%%%%%%%%%%%%%%%%%%%%%%
\newpage
%%%%%%%%%%%%%%%%%%%%%%%%%%%%%%%%%%%%%%%%%%%%%%%%%%%%%%%%%%%%%%%%%
\newpage
%%%%%%%%%%%%%%%%%%%%%%%%%%%%%%%%%%%%%%%%%%%%%%%%%%%%%%%%%%%%%%%%%
\newpage
%%%%%%%%%%%%%%%%%%%%%%%%%%%%%%%%%%%%%%%%%%%%%%%%%%%%%%%%%%%%%%%%%
\clearpage
\newpage
%%%%%%%%%%%%%%%%%%%%%%%%%%%%%%%%%%%%%%%%%%%%%%%%%%%%%%%%%%%%%%%%%
%%% NOT TO FORGIVE CHANGE CITATION INSIDE THE FIGURE
\newpage
%%%%%%%%%%%%%%%%%%%%%%%%%%%%%%%%%%%%%%%%%%%%%%%%%%%%%%%%%%%%%%%%%
%%%%%%%%%%%%%%%%%%%%%%%%%%%%%%%%%%%%%%%%%%%%%%%%%%%%%%%%%%%%%%%%%
%%% NOT TO FORGIVE CHANGE CITATION INSIDE THE FIGURE
\newpage
%%%%%%%%%%%%%%%%%%%%%%%%%%%%%%%%%%%%%%%%%%%%%%%%%%%%%%%%%%%%%%%%%
%%%%%%%%%%%%%%%%%%%%%%%%%%%%%%%%%%%%%%%%%%%%%%%%%%%%%%%%%%%%%%%%%
%%% NOT TO FORGIVE CHANGE CITATION INSIDE THE FIGURE
\newpage
%%%%%%%%%%%%%%%%%%%%%%%%%%%%%%%%%%%%%%%%%%%%%%%%%%%%%%%%%%%%%%%%%
\end{document}